# Trust-Based Mechanisms for Robust and Efficient Task Allocation in the Presence of Execution Uncertainty

**Sarvapali D. Ramchurn**                     SDR@ECS.SOTON.AC.UK
*Intelligence, Agents, Multimedia*
*School of Electronics and Computer Science*
*University of Southampton, Southampton, UK*

**Claudio Mezzetti**                          C.MEZZETTI@WARWICK.AC.UK
*Department of Economics*
*University of Warwick, Coventry, UK*

**Andrea Giovannucci**                        AGIOVANNUCCI@IUA.UPF.EDU
*SPECS Laboratory*
*Pompeu Fabra University*
*Barcelona, Spain*

**Juan A. Rodriguez-Aguilar**                 JAR@IIIA.CSIC.ES
*Artificial Intelligence Research Institute*
*Spanish Council for Scientific Research*
*Barcelona, Spain*

**Rajdeep K. Dash**                           RKD@ECS.SOTON.AC.UK
**Nicholas R. Jennings**                      NRJ@ECS.SOTON.AC.UK
*Intelligence, Agents, Multimedia*
*School of Electronics and Computer Science*
*University of Southampton, Southampton, UK*

## Abstract

Vickrey-Clarke-Groves (VCG) mechanisms are often used to allocate tasks to selfish and rational agents. VCG mechanisms are incentive compatible, direct mechanisms that are efficient (i.e., maximise social utility) and individually rational (i.e., agents prefer to join rather than opt out). However, an important assumption of these mechanisms is that the agents will *always* successfully complete their allocated tasks. Clearly, this assumption is unrealistic in many real-world applications, where agents can, and often do, fail in their endeavours. Moreover, whether an agent is deemed to have failed may be perceived differently by different agents. Such subjective perceptions about an agent's probability of succeeding at a given task are often captured and reasoned about using the notion of *trust*. Given this background, in this paper we investigate the design of novel mechanisms that take into account the trust between agents when allocating tasks.

Specifically, we develop a new class of mechanisms, called *trust-based mechanisms*, that can take into account multiple subjective measures of the probability of an agent succeeding at a given task and produce allocations that maximise social utility, whilst ensuring that no agent obtains a negative utility. We then show that such mechanisms pose a challenging new combinatorial optimisation problem (that is NP-complete), devise a novel representation for solving the problem, and develop an effective integer programming solution (that can solve instances with about $2 \times 10^5$ possible allocations in 40 seconds).

# 1. Introduction

Task allocation is an important and challenging problem within the field of multi-agent systems. The problem involves deciding how to assign a number of tasks to a set of agents according to some allocation protocol. For example, a number of computational jobs may need to be allocated to agents that run high performance computing data centres (Byde, 2006), a number of network maintenance tasks may need to be performed by communications companies for a number of business clients (Jennings, Faratin, Norman, O'Brien, Odgers, & Alty, 2000), or a number of transportation tasks may need to be allocated to a number of delivery companies (Sandholm, 1993). In the general case, the agents performing these jobs or asking for these jobs to be performed will be trying to maximise their own gains (e.g., companies owning data centres or servers will be trying to minimise the number of servers utilised, communications companies will try to minimise the number of people needed to complete the tasks demanded, and transportation companies will try to use the minimum number of vehicles). Given this, Mechanism Design (MD) techniques can be employed to design these task allocation protocols since these techniques can produce solutions that have provable and desirable properties when faced with autonomous and utility maximising actors (Dash, Parkes, & Jennings, 2003). In particular, the Vickrey-Clarke-Groves (VCG) class of mechanisms has been advocated in a number of problem domains (Walsh & Wellman, 1998; Hershberger & Suri, 2001; Dash et al., 2003) because they maximise social welfare (i.e., they are efficient) and guarantee a non-negative utility to the participating agents (i.e., they are individually rational). In such mechanisms, agents typically reveal their costs for performing the tasks or their valuation of the requested tasks to a centre and the centre then computes the allocation of tasks to each agent and the payments they all need to make and receive. However, an important underpinning assumption that such mechanisms make is that an agent *always* successfully completes every task that is assigned to it by the centre. The result of this assumption is that an allocation (i.e., an assignment of tasks that are asked for by requester agents and executed by task performer agents) is selected by the centre based only on the costs or valuations provided by the agents. This ensures that the centre always chooses the performers that are the cheapest and the requesters that are ready to pay the most. However, the agents chosen by the centre may ultimately not be successful in completing their assignment. For example, an agent providing access to a data centre with a cost of £10, but with a success rate of 100%, might be preferable to one providing the same service with a cheaper cost of £5 but with a 10% chance of being successful. Thus, in order to make efficient allocations in such circumstances, we need to design mechanisms that consider both the task performers' costs for the service *and* their *probability of success* (POS). Now, this probability may be perceived differently by different agents because they typically have different standards or means of evaluating the performance of their counterparts. For example, different customers might evaluate the performance of a data centre in different ways such as timeliness, security, or quality of the output. Given this, we turn to the notion of *trust* to capture such subjective perceptions (Ramchurn, Huynh, & Jennings, 2004). To take into account the agents' trust in other agents, as well as their costs, when allocating tasks requires the design of a new class of mechanisms that we have previously termed trust-based (Dash, Ramchurn, & Jennings, 2004).

To date, however, existing work on trust-based mechanisms (TBMs) ignores a number of important aspects of the task allocation problem which makes them less robust to uncertainty (see Section 2 for more details). First, Porter, Ronen, Shoham, and Tennenholtz (2008) only allow POS reports to come from the task performer, rather than any other agent. This means the task requester can be misled by the task performer's opinion (even if it is truthfully revealed) since the task requester may believe, at times, that the task performer failed while the task performer believes it has succeeded. Second, in our previous work (Dash et al., 2004), we presented a trust-based mechanism that could result in inefficient allocations as agents had strong incentives to over-report their POS. Even more importantly, however, existing trust-based mechanisms completely ignore the computational cost associated with including the POS and computing the optimal allocation and payments. Thus, while previous work highlights the economic benefits, they do not specify how the new problem can be effectively represented and efficiently solved. By ignoring these issues, previous work has failed to prove that such mechanisms can actually be implemented, solved, and whether they scale up to reasonable numbers of agents.

Against this background, this paper provides economically efficient and individually rational mechanisms for scenarios in which there exists uncertainty about agents successfully completing their assigned tasks. This *execution uncertainty* can generally be modelled as follows. First, potential task performers are assessed by a task requester that uses both its individual experience of their performance and information gathered from its environment (such as reports by other agents about their performance) to construct its estimation of their POS. Often these sources are called confidence and reputation respectively (Ramchurn et al., 2004; Dasgupta, 1998), and when combined they give the notion of trust in an agent performing a particular task. This combined view of trust is used here because it is a more robust measure of POS than any single estimate (especially one originating from the task performer). This is evident from the fact that each agent is only likely to have a partial view of the performance of a task performer because it is derived from a finite subset of its interactions. For example, a task requester having ten tasks performed by an agent may benefit from the experience acquired from another requester's fifty interactions with that same agent. However, incorporating trust in the decision mechanism of the requester introduces two major issues. First, when agents use reports from other agents to build trust, it introduces the possibility of *interdependent valuations*. This means that the value that is generated by one agent in the system can be affected by another agent's report to the mechanism (Jehiel & Moldovanu, 2001; Mezzetti, 2004). This, in turn, makes it much harder than in standard VCG-based techniques to incentivise agents to reveal their private information truthfully. Second, using trust to find the optimal allocation involves a significant computational cost and we show that solving the optimisation problem of trust-based mechanisms is NP-complete.

To tackle the issue of interdependence, we build upon the work by Mezzetti (2004, 2007) to construct a novel mechanism that incentivises agents to reveal their private information. Moreover, to help combat the computational complexity generated by trust, we go on to develop a novel representation for the optimisation problem posed by trust-based mechanisms and provide an implementation based on Integer Programming (IP). Given this, we show that the main bottleneck of the mechanism lies in searching through a large set of possible allocations, but demonstrate that our IP solution can comfortably solve small and medium instances within minutes (e.g., for 6 tasks and 50 agents) or hours (e.g., for 8 tasks and 70 agents).[1] In so doing, we provide the first benchmark for algorithms that aim to solve such optimisation problems.

In more detail, this paper advances the state of the art in the following ways:

---

1. Though the time taken to find the optimal solution grows exponentially with the number of tasks, our mechanism sets the baseline performance in solving the optimisation problem posed by trust-based mechanisms.

1. We design novel TBMs that can allocate tasks when there is uncertainty about their completion. Our TBMs are non-trivial extensions to the paper by Porter et al. (2008) because they are the first to consider the reputation of a task performer within the system, in addition to its self-report. This allows us to build greater robustness into the mechanism since it takes into account the subjective perceptions of all agents (task requesters in particular) about the POS of task performers.

2. We prove that our TBMs are incentive compatible, efficient and individually rational.

3. We develop a novel representation for the optimisation problem posed by TBMs and, given this, cast the problem as a special matching problem (Berge, 1973). We show that solving the generalised version of TBMs is NP-complete and provide the first Integer Programming solution for it. This solution can solve instances of 50 agents and 6 tasks within one minute and even larger instances within hours.

The rest of the paper is structured as follows. We start by providing an overview of the related work in Section 2. We then provide the contributions listed above in a step-wise manner. First, a simple task allocation model is detailed in Section 3, where we introduce the TBM for a single requester, single task scenario. Section 4 then develops the generalised TBM for multiple requesters and multiple tasks and we prove its economic properties. Having dealt with the economic aspects, we then turn to the computational problem of implementing TBMs in Section 5. Specifically, we develop a new representation for the optimisation problem posed by the generalised TBM, study the computational costs associated with solving the problem, and provide an IP-based solution to it. Section 6 then discusses a number of broader issues related to the development of future trust-based mechanisms.

## 2. Related Work

In associating uncertainty to mechanism design, we build upon work in both areas. With regards to capturing uncertainty in multi-agent interactions, most work has focused on devising computational models of trust and reputation (see papers by Teacy, Patel, Jennings, & Luck, 2006, and Ramchurn et al., 2004, for reviews). These models mostly use statistical methods to estimate the reliability of an opponent from other agents' reports and direct interactions with the opponent. Some of these models also try to identify false or inaccurate reports by checking how closely each report matches an agent's direct experience with the opponent (Teacy et al., 2006; Jurca & Faltings, 2006). Now, while these models can help in choosing the most successful agents, they are not shown to generate efficient outcomes in any given mechanism. In contrast, in this paper we provide the means to use such models in order to do just this.

In the case of MD, there has been surprisingly little work on achieving efficient, incentive compatible and individually rational mechanisms that take into account *uncertainty* in general. The approaches adopted can be separated into work on reputation mechanisms and mechanisms for task or resource allocation. The former mainly aim at eliciting honest feedback from reputation providers. Examples of such mechanisms include papers by Dellarocas (2002), Miller, Resnick, and Zeckhauser (2005), and Jurca and Faltings (2003, 2006). In particular, Miller et al. (2005) recently developed the peer prediction model, which incentivises agents to report truthfully about their experience. Their mechanism operates by rewarding reporters according to how well their reports coincide with the experience of their peers. Specifically, it assigns scores to the distance between a given agent's report and other selected *reference* reporters' reports on a given task performer. In a similar way, Jurca and Faltings (2007) have also attempted to solve the same problem by placing more importance on the repeated presence of agents in the system in order to induce truthful reporting. However, given that they focus on eliciting honest feedback, their mechanism is silent as to what this feedback is actually used for. In particular, it cannot be employed in the task allocation scenario we study in this paper because in our case the objective is to maximise the overall utility of the society that, therefore, considers the value *and* POS of agents. For example, a car repair has a lower value than building a bridge. Hence, the feedback on the car repairer is less critical than the feeback on the bridge builder in terms of its impact on the social welfare. Interestingly, their mechanism is shown to have truth-telling as a (non-unique) Nash equilibrium and it is budget balanced, but not individually rational (see Section 6 on how these social desiderata interplay).

In terms of MD for task allocation, type uncertainty is taken into account by Bayesian mechanisms such as dAGVA (d'Aspremont & Gerard-Varet, 1979; Arrow, 1979). This considers the case when the payoffs to the agents are determined via a probability distribution of types which is common knowledge to all agents. However, this mechanism cannot deal with our problem in which there is uncertainty about task completion, and each agent has information about the POS of all other agents, but there is no common knowledge of the type distributions. Porter et al. (2008) have also considered this task allocation problem and their mechanism is the one that is most closely related to ours. However, they limit themselves to the case where agents can only report on their own POS. This is a serious drawback because it assumes the agents can measure their own POS accurately and it does not consider the case where the agents may have different perceptions on the POS (e.g., a performer believes that it performs better or worse than what the requester perceives). Moreover, they only consider a single requester setting, while the mechanisms we develop here deal with multiple tasks and multiple requesters. Thus, our mechanisms can be considered to be a two-way generalisation of theirs. First, we allow multiple reports of uncertainty that need to be fused appropriately to give a precise POS as perceived by the requester. Second, we generalise their mechanism to the case of multiple requesters where the agents can provide combinatorial valuations on multiple tasks. In our earlier work on this problem (Dash et al., 2004), we proposed a preliminary TBM where the agents could have followed the risky, but potentially profitable strategy, of over-reporting their costs or under-reporting their valuations since payments are not made according to whether they succeed or fail in the allocated task (which we do in our new mechanism). In contrast, in this work, the payment scheme ensures that such a strategy is not viable and thus this mechanism is more robust. Moreover, our previous work assumed trust functions that were monotonically increasing in POS reports and (similar to Porter et al.) did not develop the algorithms that are needed to actually solve the optimisation problem posed by a TBM. In this paper, we present a mechanism that applies to more general trust functions and also develop algorithms to solve TBMs.

Finally, our work is a case of interdependent, multidimensional allocation schemes. With interdependent payoffs, Jehiel and Moldovanu (2001) have shown that is impossible to achieve efficiency with a one-stage mechanism. Mezzetti (2004), however, has shown that it is possible to achieve efficiency with an elegant two-stage mechanism under very reasonable assumptions. Our mechanism achieves efficiency without needing two reporting stages because, in the setting we consider, payments can be contingent on whether or not tasks are successful and because agents do not derive a direct payoff from the allocation of a task to another agent or the other agents' assessments about the completion probabilities. In our setting, there exists a specific function that captures the interde-pendence that exists among the agents through their assessments of each others' POS. This function is, in our case, the agents' trust model.

## 3. Single Requester, Single Task Allocation Mechanisms

In this section, we first present the basic VCG mechanism for a simple task allocation model (a single task being requested by a single agent) where the allocated task is guaranteed to be completed (i.e., all agents' POS are equal to 1). We then briefly describe Porter et al.'s (2008) extension which considers task performers that have a privately known objective probability that they finish the assigned task. Finally, we consider the case where the POS of a task performer is a function of privately known variables held by each task performer in the system. This ensures that the choice made by the task requester is better informed (drawing data from various sources) about the POS of task performers. We show how Porter et al.'s mechanism would fail to produce the efficient allocation in such settings and then go on to provide a non-trivial extension of their model to cater for this. In so doing, we define a new trust-based mechanism for the single requester, single task scenario (as a prelude to the generalised mechanism that we will develop in the next section). We then go on to prove the economic properties of this simple TBM. Throughout this section, a running example task allocation problem is employed to demonstrate the workings of the mechanisms discussed.

### 3.1 Allocation with Guaranteed Task Completion

In this task allocation scenario, a single agent derives a value when a certain task is performed. To this end, that agent needs to allocate the task to one of the available task performers, which will charge a certain amount to execute the task. We start by considering the following simple example:

**Example 1.** *MoviePictures.com, a computer graphics company, has an image rendering task that it wishes to complete for a new movie. Hence, MoviePictures.com publicly announces its intention to all companies owning data centres that can execute the task. Given the interest shown by many of these companies, MoviePictures.com needs to decide on the mechanism to allocate the contract and how much to pay the chosen contractor, given that MoviePictures.com does not know all the contractors' costs to execute the job (i.e., it does not know how much it actually costs each company to process the images and render them to the required quality).*

The above example can be captured by the following model. There is a set of agents (data centre agents in the example), $\mathcal{I} = \{1, 2, \ldots, i, \ldots, I\}$, who each have a privately-known cost $c_i(\tau) \in \mathbb{R}^+ \cup \{0\}$ of performing the rendering task $\tau$. Furthermore, let MoviePictures.com be represented by a special agent 0, who has a value $v_0(\tau) \in \mathbb{R}^+ \cup \{0\}$ for the rendering task and a cost of $c_0(\tau) > v_0(\tau)$ to perform the task ($c_0(\tau) = \infty$ in case agent 0 cannot execute the task). Hence, MoviePictures.com can only get the task performed by another agent in the set $I$ who has a cost $c_i(\tau) \le v_0(\tau)$.

Now, MoviePictures.com needs to decide on the procedure to award the contract, and hence, acts as the *centre* that will invite offers from the other agents to perform the task. In devising such a mechanism for task allocation, we focus on *incentive-compatible direct revelation* mechanisms (DRMs) by invoking the *revelation principle* which states that any mechanism can be transformed into a DRM (Krishna, 2002). In this context, "direct revelation" means the strategy space (i.e., all possible actions) of the agents is restricted to reporting their *type* (i.e., their private information, for example their cost or valuation of a task) and "incentive-compatible" means the equilibrium strategy (i.e., best strategy under a certain equilibrium concept) is truth-telling.

Thus, in a DRM, the designer has control over two parts: 1) the allocation rule that determines who wins the contract, and 2) the payment rule that determines the transfer of money between the centre (i.e., MoviePictures.com) and the agents (i.e., the data centres). Let $K$ denote a particular allocation within the space of possible allocations $\mathcal{K}$ and $\tau^{i \leftarrow 0}$ represent that agent $i$ gets allocated task $\tau$ from agent 0. Then, in this setting, the space of all possible allocations are $\mathcal{K} = \{\varnothing, \tau^{1 \leftarrow 0}, \tau^{2 \leftarrow 0}, \ldots, \tau^{I \leftarrow 0}\}$ where $\varnothing$ denotes the case where the task is not allocated. Moreover, we abuse notation slightly to define the cost of an allocation $K$ to agent $i$, as being $c_i(K) = c_i(\tau)$ if $K = \tau^{i \leftarrow 0}$ and $c_i(K) = 0$ otherwise. Similarly, for the centre, the value of a non empty allocation is simply the value it has for the task, i.e., $v_0(K) = v_0(\tau)$ if $K \neq \emptyset$ and $v_0(K) = 0$ if $K = \emptyset$. Finally, let $r_i(\cdot) \in \mathbb{R}$ be the payment by the centre to agent $i$. In case $r_i(\cdot)$ is negative, agent $i$ has to pay $|r_i(\cdot)|$ to the centre.

Within the context of task allocation, direct mechanisms take the form of sealed-bid auctions where task performers report their costs to a centre (or auctioneer). Agents may not wish to report their true costs if reporting these falsely leads to a preferable outcome for them. We will therefore distinguish between the actual costs and the reported ones by superscripting the latter with '$\,\widehat{\phantom{c}}\,$'.

The task allocation problem then consists of choosing the allocation and payment rules such that certain desirable system objectives (some of which are detailed below) are satisfied. An allocation rule is a mapping from reported costs to the set of allocations, with $K(\widehat{c}_i, \widehat{c}_{-i})$ being the allocation chosen when agent $i$ reports $\widehat{c}_i$ and all other agents report the vector $\widehat{c}_{-i}$. Similarly, a payment rule is a mapping from reported costs to payments for each agent, with $r_i(\widehat{c}_i, \widehat{c}_{-i})$ being the payment to agent $i$ when agent $i$ reports $\widehat{c}_i$ and all other agents report the vector $\widehat{c}_{-i}$.

Following the task execution and payments, an agent $i$ derives a utility given by its utility function $u_i : \mathcal{K} \times \mathbb{R} \to \mathbb{R}$. As is common in this domain, we assume that an agent is rational (expected utility maximiser) and has a quasi-linear utility function (MasColell, Whinston, & Green, 1995):

**Definition 1.** *A **quasi-linear utility function** is one that can be expressed as:*

$$u_i(K, r_i) = r_i - c_i(K) \tag{1}$$

*where $K \in \mathcal{K}$ is a given allocation.*

Having modelled the problem as above, MoviePictures.com would like to use a protocol that possesses the desirable properties of efficiency and individual rationality. It also needs to make sure that the protocol is incentive compatible: agents must find it optimal to report their true costs. These desiderata can be formally defined as follows:

**Definition 2.** ***Efficiency:*** *the allocation mechanism is said to achieve efficiency if the outcome it generates maximises the total utility of all the agents in the system (without considering transfers). That is, for all vectors of reports $\widehat{c}$, it calculates $K^*$ such that:*

$$K^*(\widehat{c}) = \arg\max_{K \in \mathcal{K}} \left[ v_0(K) - \sum_{i \in \mathcal{I}} \widehat{c}_i(K) \right] \tag{2}$$

**Definition 3.** ***Individual Rationality:*** *the allocation mechanism is said to achieve individual rationality if agents derive higher utility when participating in the mechanism than when opting out of it.*

*Assuming that the utility that an agent obtains when opting out is zero, then an individually rational allocation $K$ is one in which (Krishna, 2002):*

$$u_i(K, r_i) \geq 0 \ , \ \forall i \in \mathcal{I} \tag{3}$$

**Definition 4.** *Incentive compatibility: the allocation mechanism is said to achieve incentive compatibility if an agent's true type is its optimal report no matter what other agents report. That is:*

$$r_i(c_i, \widehat{\boldsymbol{c}}_{-i}) - c_i(K(c_i, \widehat{\boldsymbol{c}}_{-i})) \geq r_i(\widehat{c}_i, \widehat{\boldsymbol{c}}_{-i}) - c_i(K(\widehat{c}_i, \widehat{\boldsymbol{c}}_{-i})) \ \forall c_i, \forall \widehat{c}_i, \forall \widehat{\boldsymbol{c}}_{-i}.$$

Note that incentive compatibility implies that for each vector of reports of the other agents $\widehat{\boldsymbol{c}}_{-i}$ the payments to agent $i$ must depend on $i$'s own report only through the chosen allocation. Incentive compatibility requires that telling the truth be a (weakly) dominant strategy. It is also important to note that incentive compatibility in dominant strategies is the strongest possible form of incentive compatibility. The VCG mechanism has this property.

MoviePictures.com then decides to employ a Vickrey auction (also known as a second-price sealed bid auction) since this protocol possesses the desired properties of incentive compatibility, efficiency, and individual rationality (Krishna, 2002). In more detail, after having received the sealed bids (reports $\widehat{\boldsymbol{c}}$) from all the agents, the centre calculates the allocation $K^*(\widehat{\boldsymbol{c}})$ according to Equation (2), while the transfer $r_i(\cdot)$ to the winner $i$ is given by:

$$r_i(\widehat{\boldsymbol{c}}) = v_0(K^*(\widehat{\boldsymbol{c}})) - \max_{K' \in \mathcal{K}_{-i}} \left[ v_0(K') - \sum_{j \in \mathcal{I} \setminus i} \widehat{c}_j(K') \right] \tag{4}$$

where $\mathcal{K}_{-i}$ is the set of all allocations that do not involve $i$ as a task performer.

### 3.2 Allocation with Execution Uncertainty

In the mechanism presented in the previous section, it is assumed that once the allocation $K^*$ is decided, its value $v_0(K^*)$ will be obtained by the centre (either $v_0(\tau)$ if the task has been allocated or 0 otherwise). Thus, there is an implicit assumption that once allocated a task, an agent will *always* perform it successfully. However, this is unrealistic, as illustrated by the following example:

**Example 2.** *Many of the previous rendering tasks required by MoviePictures.com were allocated to PoorRender Ltd because of its very competitive prices. Unfortunately, PoorRender Ltd could not complete the task in many cases because of lack of staff and other technical problems (which it knew about before even bidding for the task). As a result, MoviePictures.com incurred severe losses. Hence, MoviePictures.com decides to alter the allocation mechanism in such a way that the agents' POS in completing the tasks can be factored into the selection of the cheapest agent. MoviePictures.com assumes each contractor knows its own POS and cost privately and needs the mechanism to elicit this information truthfully in order to choose the best allocation.*

The above problem was studied by Porter et al. (2008) and we briefly describe, in our own terms, their mechanism in order to extend and generalise it later (see Sections 3.3 and 4). We first introduce the boolean indicator variable $\kappa$ that will denote whether the task has been completed ($\kappa = 1$) or not ($\kappa = 0$). Thus, $\kappa$ is only observable after the task has been allocated. Moreover, we extend our notation here to capture the centre's valuation of the task execution such that $v_0(\kappa) = v_0(K^*)$ if $\kappa = 1$ and $v_0(\kappa) = 0$ if $\kappa = 0$. In this setting, we assume that $\kappa$ is commonly observed (i.e., if agent $i$ believes that $\kappa = 1$, then all agents $i \in \mathcal{I} \cup \{0\}$ believe the same). In our rendering example, $\kappa$ might denote whether the images are rendered up to the appropriate resolution which will allow its usage or not. Furthermore, the probability that $\kappa = 1$ once the task is allocated to agent $i$ is dependent upon another privately known variable, $p_i(\tau) \in [0,1]$, which is the POS of agent $i$ in executing task $\tau$. Note that this variable is privately known to the task performer $i$ itself, and so there is a single observation within the system, carried out by the task performer, about its own POS. Also note that the task performer incurs the cost $c_i(\tau)$ as soon as it attempts the task $\tau$ and irrespective of whether it is successful or not.

As can be seen, the value that the centre (MoviePictures.com) will derive, $v_0(\kappa)$, is not known *before* the allocation is calculated. Hence, the notions of efficiency and individual rationality introduced in section 3.1 need to be adjusted to this new setting. Given the probability that the task will be executed by a given agent, we have to consider the *expected* value of an allocation, $\overline{v}_0(K, \boldsymbol{p})$, which is calculated as:

$$\overline{v}_0(K, \boldsymbol{p}) = v_0(K) \cdot p_i(\tau) \tag{5}$$

where $i$ is the agent chosen to perform the task in allocation $K$ and $\boldsymbol{p} = \langle p_1(\tau), \ldots, p_I(\tau) \rangle$ is the vector of POS values of all the agents (the list of assessments by each contractor of its own probability that it will complete the rendering task as in our example). We now need to require agents to report their POS, in addition to the cost. We denote as $\widehat{\boldsymbol{p}}$ the vector of *reported* POS values $\langle \widehat{p}_1(\tau), \ldots, \widehat{p}_I(\tau) \rangle$.

The following modified desiderata need to be considered now:

**Definition 5.** *Efficiency: a mechanism is said to achieve efficiency if it chooses the allocation that maximises the sum of expected utilities (without considering the transfers):*

$$K^*(\widehat{\boldsymbol{c}}, \widehat{\boldsymbol{p}}) = \arg\max_{K \in \mathcal{K}} \left[ \overline{v}_0(K, \widehat{\boldsymbol{p}}) - \sum_{i \in \mathcal{I}} \widehat{c}_i(K) \right] \tag{6}$$

Note here that both $\widehat{c}_i(K)$ and $\widehat{\boldsymbol{p}}$ are reported by the agents and are key to computing the efficient allocation.

**Definition 6.** *Individual Rationality: a mechanism achieves individual rationality if a participating agent $i$ derives an expected utility, $\overline{u}_i$, which is always non-negative:*

$$\overline{u}_i(\boldsymbol{c}, \boldsymbol{p}) = \overline{r}_i(\boldsymbol{c}, \boldsymbol{p}) - c_i(K) \geq 0$$

where $\overline{r}_i(\boldsymbol{c}, \boldsymbol{p})$ is the expected payment that agent $i$ receives.

In order to achieve these desiderata, one could suppose that a naïve extension of the standard Vickrey mechanism presented above would be sufficient. In such a mechanism, the centre would ask the agents to report their extended types $(\widehat{c}_i, \widehat{p}_i(\tau))$. The allocation chosen would then be the one maximising the expected utility of the agents and the payment rule would be conditioned according to Equation (4) with $\overline{v}_0(K^*, \boldsymbol{p})$ replacing $v_0(K^*)$. However, such a mechanism would fail in these settings, as illustrated in the next section.

### 3.2.1 Naïve Application of the Vickrey Auction

**Example 3.** *Consider the case where MoviePictures.com derives a value of $v_0(\tau) = 300$ when the rendering task is completed and let there be three contractors whose costs $c_i(\tau)$ to render the images are given by $(c_1(\tau), c_2(\tau), c_3(\tau)) = (100, 150, 200)$. Furthermore, assume each contractor has a POS given by $(p_1(\tau), p_2(\tau), p_3(\tau)) = (0.5, 0.9, 1)$. This information is represented in Table 1.*

The efficient allocation in this case (shaded line in Table 1) involves assigning the task to agent 2 with an expected social utility of $300 \times 0.9 - 150 = 120$. The payment to agent 2 using the (reverse) Vickrey auction with expected values is $300 \times 0.9 - (300 - 200) = 170$ (from Equation (4)). However, such a mechanism is not incentive-compatible. For example, if agent 1 reveals that $\widehat{p_1}(\tau) = 1$, then the centre will implement $K^* = \tau^{1 \leftarrow 0}$ and will pay agent 1, $r_1 = 300 - 120 = 180$. Thus, the agents in such a mechanism are always better off reporting $\widehat{p_i}(\tau) = 1$, no matter what their actual POS is! Hence, the centre will not be able to implement the efficient allocation.

| Agent | $c_i(\tau)$ | $p_i(\tau)$ |
|-------|-------------|-------------|
| 1 | 100 | 0.5 |
| 2 | 150 | 0.9 |
| 3 | 200 | 1 |

*Table 1:* Costs of performing task $\tau$ and each agent's own perceived probability of successfully completing the task.

This type extension (i.e., including the POS) is non-trivial because the POS report of an agent affects the social value expected by the centre, but not the agent's cost under an allocation. As a result, reporting a higher POS will only positively affect an agent's probability of winning the allocation and thus will positively affect its utility. To rectify this, we need a means by which this gain in utility is balanced by a penalty so that only on truthfully reporting its type, will an agent maximise its utility. This is achieved in Porter et al.'s (2008) mechanism, which we briefly detail in the next section.

### 3.2.2 Porter et al.'s Mechanism

This mechanism is based around payments being applied *after* the completion of tasks. Specifically, the mechanism finds the marginal contribution that an agent has made to the expected welfare of other agents depending on whether it completes its assigned task or not. Intuitively, this works since the payment scheme punishes an agent that is assigned a task but does not complete it (i.e., $\kappa = 0$). As a result, the agent is not incentivised to reveal a higher POS value than its real POS since if it is then allocated the task, it is more likely to reap a punishment rather than the reward which it obtains when it successfully completes the task (i.e., $\kappa = 1$).

In more detail, the allocation is determined by the centre according to Equation (6). The payment rule for an agent $i$ to which the task $\tau$ is allocated is similar to that of the VCG in that the marginal contribution of the agent to the system is extracted by comparing the efficient allocation with the second best allocation, excluding the agent (the agent gets $r_i(\widehat{c}, \widehat{p}, \kappa) = 0$ if it is not allocated the task). The difference is that it is the *expected* marginal contribution that is extracted (i.e., taking into account the agent's real probability of success). This is achieved as follows:

$$
r_i(\widehat{\boldsymbol{c}}, \widehat{\boldsymbol{p}}, \kappa) = 
\begin{cases}
v_0(K^*(\widehat{\boldsymbol{c}}, \widehat{\boldsymbol{p}})) - \max\limits_{K' \in \mathcal{K}_{-i}} \left( \overline{v}_0(K', \widehat{\boldsymbol{p}}) - \sum_{j \in \mathcal{I} \setminus i} \widehat{c}_j(K') \right) & \text{, if } \kappa = 1 \\[4mm]
- \max\limits_{K' \in \mathcal{K}_{-i}} \left( \overline{v}_0(K', \widehat{\boldsymbol{p}}) - \sum_{j \in \mathcal{I} \setminus i} \widehat{c}_j(K') \right) & \text{, if } \kappa = 0
\end{cases}
\tag{7}
$$

where $\mathcal{K}_{-i}$ is the set of allocations excluding agent $i$.

The mechanism would work with the example provided in Table 1 since if, for example, agent 1 reports $\widehat{p}_1(\tau) = 1$, it will then be allocated the task and will be paid $300 - 120 = 180$ with a probability of 0.5 and $-120$ with a probability of 0.5. Thus, on average, agent 1 will be paid 30 but each time it will incur a cost of 100, thereby making an expected utility of $-70$. Clearly, then, a rational agent will not overstate its POS. In fact, the incentive compatibility of this mechanism arises because an agent $i$'s expected utility, given it is allocated the task, is:

$$
\begin{aligned}
\overline{u}_i(\widehat{\boldsymbol{c}}, \widehat{\boldsymbol{p}}) = {} & p_i(\tau) \left[ v_0(K^*(\widehat{\boldsymbol{c}}, \widehat{\boldsymbol{p}})) - c_i(K^*(\widehat{\boldsymbol{c}}, \widehat{\boldsymbol{p}})) - \max_{K' \in \mathcal{K}_{-i}} \left( \overline{v}_0(K', \widehat{\boldsymbol{p}}) - \sum_{j \in \mathcal{I} \setminus i} \widehat{c}_j(K') \right) \right] \\
& + (1 - p_i(\tau)) \left[ -c_i(K^*(\widehat{\boldsymbol{c}}, \widehat{\boldsymbol{p}})) - \max_{K' \in \mathcal{K}_{-i}} \left( \overline{v}_0(K', \widehat{\boldsymbol{p}}) - \sum_{j \in \mathcal{I} \setminus i} \widehat{c}_j(K') \right) \right] \\
= {} & \overline{v}_0(K^*(\widehat{\boldsymbol{c}}, \widehat{\boldsymbol{p}}), \boldsymbol{p}) - c_i(K^*(\widehat{\boldsymbol{c}}, \widehat{\boldsymbol{p}})) - \max_{K' \in \mathcal{K}_{-i}} \left( \overline{v}_0(K', \widehat{\boldsymbol{p}}) - \sum_{j \in \mathcal{I} \setminus i} \widehat{c}_j(K') \right)
\end{aligned}
\tag{8}
$$

Note that the expected utility within this mechanism is the same as what would have been derived by agents in the naïve extension of the VCG if they were truthful in reporting $\boldsymbol{p}$. However, in Porter et al.'s mechanism, agents do not have an incentive to lie. This is because, if $\widehat{p}_i(\tau) > p_i(\tau)$ (i.e., the agent over-reports its POS), then the agent might be allocated the task even though:

$$
i \neq \arg \max_{x \in \mathcal{I}} \left[ v_0(K^x) p_x(\tau) - c_x(K^x) \right]
$$

where $K^x = \tau^{x \leftarrow 0}$, which means it could be that:

$$
\overline{v}_0(K^*(\widehat{\boldsymbol{c}}, \widehat{\boldsymbol{p}}), \boldsymbol{p}) - c_i(K^*(\widehat{\boldsymbol{c}}, \widehat{\boldsymbol{p}})) < \max_{K' \in \mathcal{K}_{-i}} \left( \overline{v}_0(K', \widehat{\boldsymbol{p}}) - \sum_{j \in \mathcal{I} \setminus i} \widehat{c}_j(K') \right)
$$

This results in the agent deriving a negative utility as per Equation (8). Hence, an agent will not report higher POS values. A more complete treatment of the proof of the incentive-compatibility of the mechanism is given in the paper by Porter et al. (2008). Furthermore, the mechanism is also proven to be individually rational and efficient.

### 3.3 Allocation with Multiple Reports of Execution Uncertainty

In the previous section, we considered a mechanism in which each agent has only its privately known estimation of its own uncertainty in task completion. This mechanism considers that the centre can only receive a *single* estimate of each agent's POS. We now turn our attention to the previously unconsidered, but more general, case where *several* agents may have such an estimate. For example, a number of agents may have interacted with a given data centre provisioning company on many occasions in the past and therefore acquired a partial view on the POS of that company. Using such estimates, the centre can obtain a more accurate picture of a given agent's likely performance if it combines these different estimates together. This combination results in a better estimate for a number of reasons, including:

1. Accuracy of estimation: The accuracy of an estimation is typically affected by noise. Thus, combining a number of observations should lead to a more refined estimate than obtaining a single point estimate.

2. Personal Preferences: Each agent within the system may have different opinions as to what constitutes success when attempting a task. As a result, the centre may be willing to assign more weight to an agent's estimate if it believes this agent's perspective is more similar to its own.

We illustrate the above points by considering the following example:

**Example 4.** *MoviePictures.com is still not satisfied with the solution chosen so far. This is because PoorRender Ltd still reports that it has a high POS, even though MoviePictures.com has noticed that they have failed their task on a number of occasions. This is because PoorRender Ltd believes the images it rendered were of a high enough quality to be used in a feature film while MoviePictures.com believed they were not. MoviePictures.com therefore cannot rely on the agents' own perception of their POS to decide on the allocation. Rather, MoviePictures.com wants to ask all agents to submit their perception about each others' POS. In so doing, MoviePictures.com aims to capture the knowledge that agents might have about each other either from previous sub-contracted tasks or simple observations. To this end, MoviePictures.com needs to devise a mechanism that will capture all the agents' perceptions (including its own) into measures of POS for each agent and use these fused measures in the selection process.*

The above example can be modelled by introducing a new variable, the Expected Quality of Service (EQOS), noted as $\eta_i^j(\tau)$, which is the perception of each agent $i$ about the POS of agent $j$ on task $\tau$. Now, the vector of agent $i$'s EQOS of all agents (including itself) within the system is noted as $\boldsymbol{\eta}_i = \langle \eta_i^1(\tau), \ldots, \eta_i^I(\tau) \rangle$. Furthermore, we shall denote as $\boldsymbol{\eta}^j$ the EQOS that all agents within the system (including itself) have about agent $j$. Thus, in our image rendering example, $\eta_i^j(\tau)$ might denote the probability as perceived by agent $i$ that the rendering task is completed according to a certain level of quality of the computer graphics (which is perceived differently by the different agents). Then, MoviePictures.com needs a function in order to combine the EQOS of all the agents so as to give it a resultant POS that the movie is rendered up to its own graphic requirements.

In more detail, given $i$'s previous personal interaction with $j$, $i$ can compute, based on the frequency of good and bad interactions, a probability, termed its *confidence*, in $j$ as the POS. Second, $i$ can also take into account other agents' $(-i)$ opinions about $j$, known as $j$'s *reputation* in the society, in order to compute the POS of $j$ (Ramchurn et al., 2004). The combination of both measures is generally captured by the concept of trust, which is defined as the aggregate expectation, derived from the history of direct interactions and information from other sources, that $j$ will complete the task assigned to it. The aggregate trust that agent $j$ will successfully complete task $\tau$ for agent $0$ is a function $tr_0^j : [0,1]^{|\mathcal{I}|} \to [0,1]$.

There are multiple ways in which the trust function could be computed, but it is often captured as follows:

$$tr_0^j(\boldsymbol{\eta}) = \sum_{l \in I} w_l \times \eta_l^j \tag{9}$$

where $w_l \in [0,1]$ and $\sum w_l = 1$. This function generates trust as a weighted sum of EQOS values. In some cases, the $\eta$'s are actually considered to be probability distributions and the trust function is the expected value of the joint distribution constructed from the individually reported distributions (Teacy et al., 2006; Jurca & Faltings, 2007). Much work exists in the literature that deals with different ways of combining these distributions such that biases or incompatibilities between agents' perceptions are taken into account. Essentially, however, they all assign weights to different reports of the agents and choose the expected value of these reports as the trust in an agent. However, to date, none of these models actually studies how to get self-interested agents to generate such reports truthfully along with maximising the social welfare.

Now, a direct mechanism in this case elicits from each agent $i$, its cost and EQOS vector, $\{c_i(\tau), \boldsymbol{\eta}_i\}$, after which the centre decides on the allocation and payments to the agents. In computing its expected utility in a mechanism, an agent must evaluate the trust, or probability of success, by the agent who is allocated the task. This raises a conceptual difficulty. How should an agent treat the other agents' POS reports in assessing the probability of task completion (as opposed to computing its best response to their type reports)? The approach we will take in this paper is that an agent assumes the reported POS of the other agents is truthful in computing the trust in another agent; more precisely, an agent computes the value of the trust function by using his true EQOS and the reported EQOS of the other agents. Thus, the trust of agent $i$ that agent $j$ will be able to complete the task is $tr_0^j(\boldsymbol{\eta}_i, \widehat{\boldsymbol{\eta}}_{-i})$. As we have already seen, in general a payment to an agent depends on the reported types of all agents and on whether the task succeeds or fails. To this end, let $i(\widehat{\boldsymbol{c}}, \widehat{\boldsymbol{\eta}})$ be the agent who is allocated the task when the vector of reported types is $(\widehat{\boldsymbol{c}}, \widehat{\boldsymbol{\eta}})$. Then, define the expected payment to agent $i$ when the true types are $(\boldsymbol{c}, \boldsymbol{\eta})$ and the reported types are $(\widehat{\boldsymbol{c}}, \widehat{\boldsymbol{\eta}})$ as follows:

$$Er_i(\widehat{\boldsymbol{c}}, \widehat{\boldsymbol{\eta}}; \boldsymbol{c}, \boldsymbol{\eta}) = r_i(\widehat{\boldsymbol{c}}, \widehat{\boldsymbol{\eta}}, \kappa = 1) tr_0^{i(\widehat{\boldsymbol{c}}, \widehat{\boldsymbol{\eta}})}(\boldsymbol{\eta}_i, \widehat{\boldsymbol{\eta}}_{-i}) + r_i(\widehat{\boldsymbol{c}}, \widehat{\boldsymbol{\eta}}, \kappa = 0) \left[ 1 - tr_0^{i(\widehat{\boldsymbol{c}}, \widehat{\boldsymbol{\eta}})}(\boldsymbol{\eta}_i, \widehat{\boldsymbol{\eta}}_{-i}) \right]$$

We should point out that the type of an agent (EQOS plus cost) is multidimensional and, as is common in a multidimensional world, there could be several type reports that generate the same expected payment to an agent. We are now ready to define the modified notion of incentive compatibility we will use.[2]

**Definition 7.** *Incentive compatibility (in Dominant Strategies): the allocation mechanism is said to achieve incentive compatibility in dominant strategies if an agent's true type is its optimal report no matter what the other agents report. That is:* $\forall \boldsymbol{c}, \forall \boldsymbol{\eta}, \forall \widehat{c}_i, \forall \widehat{\boldsymbol{\eta}}_i, \forall \widehat{\boldsymbol{c}}_{-i}, \forall \widehat{\boldsymbol{\eta}}_{-i}$,

$$Er_i(c_i, \widehat{\boldsymbol{c}}_{-i}, \boldsymbol{\eta}_i, \widehat{\boldsymbol{\eta}}_{-i}; \boldsymbol{c}, \boldsymbol{\eta}) - c_i(K(c_i, \widehat{\boldsymbol{c}}_{-i})) \geq Er_i(\widehat{c}_i, \widehat{\boldsymbol{c}}_{-i}, \widehat{\boldsymbol{\eta}}_i, \widehat{\boldsymbol{\eta}}_{-i}; \boldsymbol{c}, \boldsymbol{\eta}) - c_i(K(\widehat{c}_i, \widehat{\boldsymbol{c}}_{-i}))$$

---

2. That an agent uses the reported POS of the other agents in computing the value of the trust function seems a natural assumption when an agent can rely on the other agents truthfully reporting their types. This is the case, for example, when the history of interactions between the POS reporters is publicly known (e.g., on eBay or Amazon).

Now, in the case where agents do not view the EQOS reports of the other agents as being truthful, the trust of agent $i$ that agent $j$ will be able to complete the task may depend on both true and reported types of all agents; in such a case we could relax the incentive compatibility requirement from dominant strategy to (ex-post) Nash equilibrium (MasColell et al., 1995), which means that if all the other agents report truthfully, then it is optimal for an agent always to report its true type, no matter what the true types of the other agents are. After replacing the new trust function in the definition of the expected payment to agent $i$, the definition of incentive compatibility would change to:

**Definition 8.** *Incentive compatibility (in Nash Equilibrium): the allocation mechanism is said to achieve incentive compatibility in (ex-post) Nash equilibrium if an agent's true type is its optimal report provided other agents report their type truthfully. That is:* $\forall c_i, \forall \widehat{c}_i, \forall \boldsymbol{\eta}_i, \forall \widehat{\boldsymbol{\eta}}_i, \forall \boldsymbol{c}_{-i}, \forall \boldsymbol{\eta}_{-i}$,

$$Er_i(c_i, \boldsymbol{c}_{-i}, \boldsymbol{\eta}_i, \boldsymbol{\eta}_{-i}; \boldsymbol{c}, \boldsymbol{\eta}) - c_i(K(c_i, \boldsymbol{c}_{-i})) \geq Er_i(\widehat{c}_i, \boldsymbol{c}_{-i}, \widehat{\boldsymbol{\eta}}_i, \boldsymbol{\eta}_{-i}; \boldsymbol{c}, \boldsymbol{\eta}) - c_i(K(\widehat{c}_i, \boldsymbol{c}_{-i}))$$

We next demonstrate why Porter et al.'s mechanism would not work in this setting by extending example 1.

### 3.3.1 FAILURE OF PORTER ET. AL'S MECHANISM

**Example 5.** *Two agents have costs for performing a task $\tau$ requested by the centre and have formed perceptions on the set of agents $\mathcal{I}$ given in Table 2. Suppose that $tr_0^i(\boldsymbol{\eta}) = [\eta_1^i(\tau) + \eta_2^i(\tau)]/2$, and $v_0(\tau) = 1$.*

| Agent | $c_i(\tau^{i\leftarrow})$ | $\eta_i^1(\tau)$ | $\eta_i^2(\tau)$ |
|-------|------|------|------|
| 1 | 0 | 0.6 | 1 |
| 2 | 0 | 0.8 | 0.6 |
| $tr_0^i(\boldsymbol{\eta})$ | | 0.7 | 0.8 |

*Table 2:* Costs and EQOS reports of agents in a single task scenario. The trust of the requester is calculated assuming truthful reports.

Porter et al. do not specify a procedure that deals with EQOS reports. However, a natural extension of their technique would be to allocate according to $tr_0^i(\widehat{\boldsymbol{\eta}})$ instead of $\widehat{p}_i(\tau)$, and to ignore all reports of agent $i$ in the computation of its payment. We implement this in the above example. Agent 2 should be the winner since it generates an expected social utility of 0.8, while agent 1 would generate a utility of 0.7. The expected utility to the agent allocated the task is then (according to Equation (8)):

$$\overline{u}_i(\widehat{\boldsymbol{c}}, \widehat{\boldsymbol{\eta}}) = v_0(K^*(\widehat{\boldsymbol{c}}, \widehat{\boldsymbol{\eta}})) \cdot tr_0^i(\boldsymbol{\eta}_i, \widehat{\boldsymbol{\eta}}_{-i}) - c_i(K^*(\widehat{\boldsymbol{c}}, \widehat{\boldsymbol{\eta}})) - \max_{K' \in \mathcal{K}_{-i}} \left[ v_0(K') \cdot tr_0^{j'}(\widehat{\boldsymbol{\eta}}_{-i}) - \widehat{c}_j(K') \right]$$
$$(10)$$

where $\widehat{\boldsymbol{\eta}}_{-i}$ excludes all $\eta$ reports by agent $i$, $\mathcal{K}_{-i}$ is the set of allocations excluding agent $i$, and $j'$ is the agent that is allocated the task under allocation $K'$. Unfortunately, this extension breaks incentive compatibility in the following way. Given that the efficient allocation is computed using the *reported* $\widehat{\boldsymbol{\eta}}$ values of *all* agents (using $tr_0(\widehat{\boldsymbol{\eta}})$ instead of $\widehat{\boldsymbol{p}}$ in Equation (6)), the value of the best allocation obtained by removing one agent could be arbitrarily lower. In the example above, if agent 1 reports $\eta_1^2 = 0$, the efficient allocation becomes agent 1 with an expected social utility of 0.7 and agent 1 gets an expected utility of 0.1 because the system's utility drops to 0.6 when its reports are removed and the allocation recomputed. If agent 1 is truthful it will obtain 0 utility since agent 2 would be the winner in this case. In effect, the removal of an agent from the system breaks the mechanism because of the *interdependence* between the valuations introduced by the trust model. We elaborate further on this issue and show how to solve it in the next section.

We thus need to develop a mechanism that is incentive-compatible when agents are reporting about their perceptions of other agents' POS. In order to do so, however, we now need to additionally consider the effect that reporting the EQOS vector has on an agent's expected utility. Specifically, we need to develop a trust-based mechanism in which the EQOS reports of an agent do not provide it with a way of increasing its overall expected utility (as per the intuition behind the VCG). Then, with the true value of the EQOS, the mechanism will result in the selection of the optimal allocation of tasks.

### 3.3.2 The Single Requester Single Task Trust-Based Mechanism

Intuitively, the following mechanism works by ascertaining that an agent derives a positive utility when it successfully completes a task and its EQOS report does not change the allocation in its favour (thus, the mechanism we develop can be regarded as a generalisation of the paper by Porter et al., 2008).

In more detail, let $i(K)$ be the agent performing the task under allocation $K$; the centre first determines the allocation according to:

$$K^*(\widehat{\boldsymbol{c}}, \widehat{\boldsymbol{\eta}}) = \arg\max_{K \in \mathcal{K}} \left[ v_0(K) \cdot tr_0^{i(K)}(\widehat{\boldsymbol{\eta}}) - \sum_{i \in \mathcal{I}} \widehat{c}_i(K) \right] \tag{11}$$

Having computed the efficient allocation as above, we adopt a similar approach to Porter et al.'s to compute the payments after tasks have been executed (see section 3.2.2). However, the novelty of our mechanism lies in the use of *all* agents' EQOS reports in the computation of the efficient allocation (as we showed above). Moreover, we have additional payments for the losers to incentivise all agents to select the efficient allocation.

Thus, we apply different payments to the cases where the agent winning the allocation succeeds (i.e., $\kappa = 1$) and when it fails (i.e., $\kappa = 0$). So if agent $i$ is allocated the task (i.e., $K^* = \{\tau^{i \leftarrow 0}\}$) the payment is:

$$r_i(\widehat{\boldsymbol{c}}, \widehat{\boldsymbol{\eta}}, \kappa) = \begin{cases} v_0(K^*(\widehat{\boldsymbol{c}}, \widehat{\boldsymbol{\eta}})) - B_i(\widehat{\boldsymbol{c}}_{-i}, \widehat{\boldsymbol{\eta}}_{-i}) & \text{, if } \kappa = 1 \\ -B_i(\widehat{\boldsymbol{c}}_{-i}, \widehat{\boldsymbol{\eta}}_{-i}) & \text{, if } \kappa = 0 \end{cases} \tag{12}$$

where $B_i(\cdot) \geq 0$ is a term independent from $i$'s report (a constant from $i$'s point of view) that reduces the payment that needs to be made to the agent. We briefly discuss how the value of $B_i(\cdot)$ could be set to reduce the payout made by the centre later in this section, and we provide greater detail in section 4.4.

In addition to paying the winner, we also reward the losers $k \in I \setminus i$ in the following way, depending on whether $i$ succeeds or not:

$$r_k(\widehat{\boldsymbol{c}}, \widehat{\boldsymbol{\eta}}, \kappa) = \begin{cases} v_0(K^*(\widehat{\boldsymbol{c}}, \widehat{\boldsymbol{\eta}})) - \widehat{c}_i(K^*(\widehat{\boldsymbol{c}}, \widehat{\boldsymbol{\eta}})) - B_k(\widehat{\boldsymbol{c}}_{-k}, \widehat{\boldsymbol{\eta}}_{-k}) & \text{, if } \kappa = 1 \\ \\ -\widehat{c}_i(K^*(\widehat{\boldsymbol{c}}, \widehat{\boldsymbol{\eta}})) - B_k(\widehat{\boldsymbol{c}}_{-k}, \widehat{\boldsymbol{\eta}}_{-k}) & \text{, if } \kappa = 0 \end{cases} \tag{13}$$

Intuitively, the payment scheme aims to incentivise all agents to reveal their type so that the most efficient allocation is chosen. Let $K_0^i$ be the allocation assigning the task to agent $i$. Suppose agent $i$ with type $(c_i, \boldsymbol{\eta}_i)$ reports its type as $(\widehat{c}_i, \widehat{\boldsymbol{\eta}}_i)$ and all other agents report $(\widehat{\boldsymbol{c}}_{-i}, \widehat{\boldsymbol{\eta}}_{-i})$. When agent $i$ wins the task, it will derive the following expected utility:

$$\overline{u}_i\left(K_0^i, \boldsymbol{\eta}_i, \widehat{\boldsymbol{\eta}}_{-i}\right) = v_0\left(K_0^i\right) \cdot tr_0^i\left(\boldsymbol{\eta}_i, \widehat{\boldsymbol{\eta}}_{-i}\right) - c_i\left(K_0^i\right) - B_i(\widehat{\boldsymbol{c}}_{-i}, \widehat{\boldsymbol{\eta}}_{-i}) \tag{14}$$

Note that $tr_0^i\left(\boldsymbol{\eta}_i, \widehat{\boldsymbol{\eta}}_{-i}\right)$ reflects the true POS of agent $i$. When agent $k \neq i$ is assigned the task, agent $i$ obtains the following expected utility by participating in the mechanism:

$$\overline{u}_i\left(K_0^k, \boldsymbol{\eta}_i, \widehat{\boldsymbol{\eta}}_{-i}\right) = v_0\left(K_0^k\right) \cdot tr_0^k\left(\boldsymbol{\eta}_i, \widehat{\boldsymbol{\eta}}_{-i}\right) - \widehat{c}_k\left(K_0^k\right) - B_i(\widehat{\boldsymbol{c}}_{-i}, \widehat{\boldsymbol{\eta}}_{-i}) \tag{15}$$

The only difference between Equations (14) and (15) is the identity of the winner. Hence, by falsely reporting, agent $i$ can only influence the identity of the winner. Agent $i$'s expected utility in the mechanism is equal to the expected social utility in the system minus a constant independent of $i$'s report. Hence, if agent $i$ is rational it should report its true type, so that the efficient agent (outcome) is chosen. This shows that the single task trust-based mechanism is incentive compatible and efficient.[3]

**Proposition 1.** *The mechanism described by Equations* (11)*,* (12)*, and* (13) *is incentive compatible.*

**Proposition 2.** *The mechanism described by Equations* (11)*,* (12)*, and* (13) *is efficient.*

*Proof.* Since agent $k$'s report about $\boldsymbol{\eta}_k$ affects the expected utility of all other agents (see Equations (14) and (15)), we have interdependence between agents' payoffs, or valuations. However, no agent can influence its own transfer through its report, because the computation of agent $i$'s payment is independent of its report $\widehat{\boldsymbol{\eta}}_i$ (and $\widehat{c}_i$) and is only dependent on the *actual execution* of the task and therefore on the true $\boldsymbol{\eta}_i$ value. It is this feature that permits the implementation of the efficient allocation with a single-stage mechanism. □

To exemplify the payments in our mechanism, consider the following extension of Example 5.

**Example 6.** *Two agents have zero cost for performing a task $\tau$ requested by the centre and have EQOS $\eta_i^j(\tau) \in \{0.6, 0.7, 0.8\}$ for $i, j = 1, 2$. Suppose that $tr_0^i(\boldsymbol{\eta}) = [\eta_1^i(\tau) + \eta_2^i(\tau)]/2$, and $v_0(\tau) = 1$.*

By setting $B_i = 0.6$ in the above example, we have that the payment to each agent when the task is completed successfully is $0.4$, while the payment when the task fails is $-0.6$. Hence, the centre profits from implementing the mechanism. Agents have an incentive to report truthfully, so that the agent most likely to succeed is allocated the task. Furthermore, all agents are willing to participate, because the probability of success is at least $0.6$ (it is $0.6$ in the worst case scenario) and hence, agents expect to obtain at least zero from participating: the mechanism is individually rational. Also note that the total expected payment from the centre to all agents is at most $0.8 \times 0.4 \times 2 - 0.2 \times 0.6 \times 2 = 0.4$, but could be as low as $0.6 \times 0.4 \times 2 - 0.4 \times 0.6 \times 2 = 0$. As we now show, $B_i$ can always be chosen so that individual rationality is satisfied.

---

3. We provide a more detailed proof for the generalised case in Section 4.3.

**Proposition 3.** *For an appropriate choice of $B_i(\cdot)$, the mechanism described by Equations* (11)*,* (12)*, and* (13) *is individually rational.*

*Proof.* By not participating in the mechanism, an agent can only obtain 0 utility. However, if an agent decides to participate, and by virtue of the selection of the efficient allocation (which returns no allocation if the social welfare generated is less than 0), it is guaranteed, as a winner, to obtain the utility $u_i$ described in Equation (14) or, as a loser, the utility $u_k$ in Equation (15). Since in both cases $u_i \geq -B_i(\cdot)$ when the efficient allocation is chosen, and $B_i$ can be set to 0, the mechanism is individually rational. □

Obviously, since all agents' utilities are tied to that of the winning agent, they also lose out if the winning agent fails but, in expectation, all agents make a profit of at least 0 in case $B_i$ is set to 0. As Example 6 shows, if the centre is trying to minimise payments (and increase its own profits), it could set $B_i$ to be greater than zero and still satisfy individual rationality. In Section 4.4, we show how to set $B_i$ to a value that maintains individual rationality while minimising payments in the general model.

Here we note that sometimes it may be preferable for the centre to give up individual rationality. Consider, for example, if we modify Example 6 to allow for an additional EQOS value $\eta_i^j(\tau) = 0.3$ for $i, j = 1, 2$. To induce type $\eta_i^j(\tau) = 0.3$ to participate, the centre could set $B_i(\cdot) = 0.3$, so that the payment following success is 0.7 and the payment after failure is $-0.3$. In the worst case scenario for the centre (i.e., when the centre's profit is the lowest), the total expected payment in this mechanism is $0.8 \times 0.7 \times 2 - 0.2 \times 0.3 \times 2 = 1$ (in the best case scenario, the total expected payment is zero). As we shall see in Section 4.4, the centre could substantially reduce its payments by making $B_i(\cdot)$ depend on the report of the other agents (i.e., other than $i$). Still, it may be preferable for the centre to set $B_i(\cdot) = 0.6$, giving up on the participation of agents with EQOS values $\eta_i^j(\tau) = \eta_i^j(\tau) = 0.3$. In general, when there are low EQOS types, the centre faces a trade off between efficient task allocation and payments minimisation. We leave the study of this trade-off to future work (see Section 6 for some initial thoughts).

## 4. The Generalised Trust-Based Mechanism

The mechanisms we presented in the previous section dealt with the basic task allocation problem in which there is one requester, one task, and several performers. Here, we aim to efficiently solve the more general problem of trust-based interactions in which more than one agent requests or performs (or both) more than one task. To this end, we extend the single requester single task setting to the more general one of multiple requesters and multiple tasks in our *Generalised Trust-Based Mechanism* (GTBM). This extension needs to consider a number of complex features on top of those dealt with previously. First, we need to consider multiple requesters that can each make requests for sets of tasks and task performers that can each perform sets of tasks as well. Thus, the centre now acts as a clearing house, determining the allocation and payments from the multiple bids from the task requesters and multiple asks from the task performers. This significantly complicates the problem of incentivising agents to reveal their types since we now have to make sure that the agents reveal their costs, valuations, and EQOS truthfully over more than one task. Second, the computation of the efficient allocation and payments will have to consider a much larger space than previously. Thus, we believe it is important to show how the problem can be modelled, implemented, and solved to demonstrate how our mechanism scales with increasing numbers of agents and tasks (the computability aspects are dealt with in Section 5).

The following example illustrates this more general setting.

**Example 7.** *After using the trust-based mechanism for a few months, MoviePictures.com made significant profits and expanded into several independent business units, each performing rendering tasks or having rendering tasks performed for certain clients. Now, MoviePictures.com would like to find ways in which its business units can efficiently allocate tasks amongst themselves. However, some companies have uncertainties about each other's performance of the rendering tasks. For example, while some business units, such as HighDefFilms.com, believe PoorRender Ltd (now part of MoviePictures.com) is inefficient, some others, such as GoodFilms.com, believe it is not so bad, having recently had a large set of animations rendered very well for a very cheap price. To cater for these differences in opinion while maximising the overall utility, MoviePictures.com needs to extend the single task trust-based mechanism and implement the generalised mechanism efficiently.*

In order to deal with this more complex setting, we extend our task allocation model in the next subsection, before describing the allocation rule and payment scheme in Section 4.2 and proving the economic properties of the mechanism in Section 4.3.

### 4.1 The Extended Task Allocation Setting

Let $\mathcal{T} = \{\tau_1, \tau_2, ..., \tau_M\}$ denote the set of tasks which can be requested or performed (compared to the single task before). We use the notation $\boldsymbol{\tau}^{\cdot \leftarrow i}$ to specify that the subset of tasks $\boldsymbol{\tau} \subseteq \mathcal{T}$ is performed specifically for agent $i$.[4] Similarly, by adding the superscript to the task, $\boldsymbol{\tau}^{i \leftarrow \cdot} \subseteq \mathcal{K}$ denotes a subset of tasks that agent $i$ performs. Note that there is nothing in our model that restricts an agent to be *only* a task performer or requester.

A selected allocation $K$ in this multiple task, multiple requester model then generates a matching problem that involves finding agents that will perform the tasks that are requested by some other agents (e.g., $K = \{\tau_1^{1 \leftarrow 1}, \tau_1^{1 \leftarrow 2}, \ldots, \tau_1^{I-1 \leftarrow I}, \ldots, \tau_M^{I \leftarrow I}\}$). Let the set of all possible allocations be denoted as $\mathcal{K}$. Note that not all requested tasks need to be allocated: that is, the matching in $K$ need not be perfect.

In the multiple task case, agents may express valuations and costs for sets of tasks as well as subsets of these sets of tasks. For example, agent $i$ may have $v_i(\tau_1, \tau_2, \tau_3) = 100$ and $v_i(\tau_1, \tau_2) = 10$ and $v_i(\tau_3) = 0$. Then, if agent $i$ gets $\tau_1, \tau_2$ and $\tau_3$ executed it gets a value of 100, while if only $\tau_1$ and $\tau_2$ get executed and $\tau_3$ fails, agent $i$ still obtains a value of 10. Similarly, agent $i$ may have task execution costs $c_i(\tau_4, \tau_5, \tau_6) = 100$ and $c_i(\tau_4, \tau_5) = 40$ and $c_i(\tau_6) = 10$. To capture such inter-relationships between valuations, let $K_i^j$ be the set of tasks within the allocation $K$ which have to be performed by agent $j$ for agent $i$ ($K_i^j$ could be the empty set). Note that each task is specific to a task requester. This means that if agents 1 and 2 request task $\tau_m$, then a task performer (putting in one bid for $\tau_m$) matched to $\tau_m$ for agent 1, only performs it for agent 1 and not for agent 2. We will abuse notation slightly and define $K = \{K_i, K^i\}_{i \in \mathcal{I}}$ where $K_i = (K_i^1, ..., K_i^I)$ and $K^i = (K_1^i, ..., K_I^i)$. An agent $i$ has a value (assuming *all* the tasks in $K$ will be completed) and cost for an allocation $K$, $v_i(K) \in \Re^+ \cup \{0\}$ and $c_i(K) \in \Re^+ \cup \{0\}$ respectively, whereby:[5]

---

4. In this paper, we will not consider agents requesting the performance of multiple units of tasks. Although our model is easily extensible to this case, the explanation is much more intricate.

$$v_i(K) = v_i(K_i)$$

$$c_i(K) = c_i(K^i)$$

Moreover, within our model, each agent $i$ has an EQOS vector, $\boldsymbol{\eta}_i = \{\eta_i^j(K_h)\}_{j,h \in \mathcal{I}}^{K_h \subseteq \mathcal{K}}$ that represents its belief in how successful all agents within the system are at completing the tasks $K_h$ for agent $h$. Thus, at the most general level, agent $i$'s type is now given by $\theta_i = \{\boldsymbol{v}_i, \boldsymbol{c}_i, \boldsymbol{\eta}_i\}$. For any given set of tasks $K_i^j$ that $j$ must perform for $i$, for any subset of tasks $\widetilde{K}_i^j \subseteq K_i^j$ and for any EQOS vector $\boldsymbol{\eta}$, we let $tr_i^j\left(\left.\widetilde{K}_i^j\right| K_i^j, \boldsymbol{\eta}\right)$ be the trust that exactly the set of tasks $\widetilde{K}_i^j$ will be completed by $j$. The trust can be computed as we have shown in Section 3.3 by simply replacing agent 0 with agent $i$ and replacing the single task by the set of tasks $\mathcal{T}$. As in the single requester case, the trust function represents the aggregate belief that agents have about a given task performer and hence all task requesters form the same probability of success (give all agents' EQOS reports) about a given task performer. Finally, we let $tr_i\left(\left.\widetilde{K}_i\right| K_i, \boldsymbol{\eta}\right) = \prod_{j \in \mathcal{I}} tr_i^j\left(\left.\widetilde{K}_i^j\right| K_i^j, \boldsymbol{\eta}\right)$.

We are now ready to present the generalised trust-based mechanism.

## 4.2 The Allocation Rule and Payment Scheme

In our generalised mechanism (GTBM), the task requesters first provide the centre with a list of tasks they require to be performed, along with their valuation vector associated with each set of tasks, whereas the task performers provide their costs for performing sets of tasks.[6] All agents also submit their EQOS vector to the centre. Thus, each agent provides the centre with reports $\widehat{\boldsymbol{\theta}}_i = \{\widehat{\boldsymbol{v}}_i, \widehat{\boldsymbol{c}}_i, \widehat{\boldsymbol{\eta}}_i\}$, so that $\widehat{\boldsymbol{\theta}} = (\widehat{\boldsymbol{\theta}}_1, ..., \widehat{\boldsymbol{\theta}}_I)$ is the report profile. Given this, the centre applies the rules of the mechanism in order to find the allocation $K^*$ and net payments $r_i$ to each agent $i$. In more detail:

1. The centre computes the allocation according to the following:

$$K^*\left(\widehat{\boldsymbol{\theta}}\right) = \underset{K = \{K_i, K^i\}_{i \in \mathcal{I}} \in \mathcal{K}}{\arg\max} \sum_{i \in \mathcal{I}} \left[ \sum_{\widetilde{K}_i \subseteq K_i} \widehat{v}_i(\widetilde{K}_i) \cdot tr_i\left(\left.\widetilde{K}_i\right| K_i, \widehat{\boldsymbol{\eta}}\right) - \widehat{c}_i(K) \right] \qquad (16)$$

   Thus, the centre uses the reports of the agents in order to find the allocation that maximises the expected utility of all agents within the system.

2. The agents carry out the tasks allocated to them in the allocation vector $K^*\left(\widehat{\boldsymbol{\theta}}\right)$.

---

5. As a result of this setup, an agent $i$ may not want some sets of tasks to be performed or it may be unable to perform such tasks. In such cases, we then assign a default value of 0 and cost of $\infty$ to those sets of tasks.

6. As noted before, task performers can also be task requesters at the same time (and vice versa).

3. The centre computes the payments to the agents, conditional on completion of the tasks allocated. Let $\kappa(K_i)$ be an indicator function that takes the value one if $K_i$ is the set of all the tasks (requested by agent $i$ from all agents) that are completed, and takes the value of zero otherwise. The payment to agent $i$ is as follows:

$$r_i\left(\widehat{\boldsymbol{\theta}}, \kappa(\cdot)\right) = \sum_{j \in \mathcal{I} \setminus i}\left[\sum_{\widetilde{K}_j \subseteq K_j^*(\widehat{\theta})} \widehat{v}_j\left(\widetilde{K}_j\right) \cdot \kappa\left(\widetilde{K}_j\right) - \widehat{c}_j\left(K^*\left(\widehat{\boldsymbol{\theta}}\right)\right)\right] - B_i(\widehat{\boldsymbol{\theta}}_{-i}) \quad (17)$$

where $B_i(\widehat{\boldsymbol{\theta}}_{-i}) \geq 0$ is a constant from $i$'s point of view (i.e., it is computed independently of agent $i$'s reports, but it may depend on the reports of the other agents), that can be used to reduce the payout that the centre has to make.

As we discussed in Section 3.3.2, the centre faces a trade-off. By reducing the value of $B_i(\cdot)$ it induces participation by a larger set of types (i.e., types with low EQOS), but it increases the centre's payments to agents, making the mechanism less profitable for the centre. Thus, the scale of the payments one might expect from application of the GTBM depends on whether the centre decides to satisfy the individual rationality constraint, thus making sure that every type wants to participate. As we shall see in Section 4.4, if the centre decides to satisfy the individual rationality constraint, then the scale of payments to agent $i$ increases with the lower bound on trust values that could be derived using $i$'s EQOS report.

It should also be noted that the computation of the payments requires solving several optimisation problems (i.e., finding the optimal allocation with and without several reports). As the number of agents increases, the difficulty of computing payments will increase and it is important to show how such payments can be efficiently computed. We elaborate on our solution to this in Section 5. Before doing so, however, we detail and prove the economic properties of our mechanism in what follows.

### 4.3 Economic Properties

Here, we provide the proofs of the incentive compatibility[7] and efficiency of the mechanism. We also prove that there are values of $B_i$ which make the mechanism individually rational.

**Proposition 4.** *The GTBM is incentive compatible.*

*Proof.* In order to prove incentive-compatibility, we will analyse agent $i$'s best response (i.e., its best report of $\widehat{\boldsymbol{\theta}}_i = \{\widehat{v}_i, \widehat{c}_i, \widehat{\boldsymbol{\eta}}_i\}$) when all other agents report $\widehat{\boldsymbol{\theta}}_{-i}$. We first calculate the expected utility that an agent $i$ will derive given the above mechanism.

---

7. Again, we place the same caveat on the notion of incentive compatibility we use here as we do in in Section 3.3 (i.e., Dominant Strategy or (ex-post) Nash equilibrium depending on whether an agent computes the trust functions by using the other agents' POS reports as if they were true or not).

The expected utility of an agent $i$ when the reported types are $\widehat{\boldsymbol{\theta}}$ and the true types are $\boldsymbol{\theta}$ is given by:

$$
\begin{aligned}
\overline{u}_i\left(\widehat{\boldsymbol{\theta}}; \boldsymbol{\theta}\right) = & \sum_{\widetilde{K}_i \subseteq K_i^*\left(\widehat{\boldsymbol{\theta}}_i, \widehat{\boldsymbol{\theta}}_{-i}\right)} v_i(\widetilde{K}_i) \cdot tr_i\left(\widetilde{K}_i \big| K_i^*\left(\widehat{\boldsymbol{\theta}}_i, \widehat{\boldsymbol{\theta}}_{-i}\right), \boldsymbol{\eta}_i, \widehat{\boldsymbol{\eta}}_{-i}\right) \\
& - c_i\left(K^*\left(\widehat{\boldsymbol{\theta}}_i, \widehat{\boldsymbol{\theta}}_{-i}\right)\right) + Er_i\left(\widehat{\boldsymbol{\theta}}; \boldsymbol{\theta}\right)
\end{aligned}
\tag{18}
$$

where $Er_i$ is the expectation of $r_i$ taken with respect to the likelihood of task completion. The probability attached by $i$ to the indicator variable $\kappa(\widetilde{K}_j)$ being equal to one (i.e., all tasks $\widetilde{K}_j$ being completed), given that the set of tasks requested by $j$ is $K_j$ and all agents different from $i$ report $\widehat{\boldsymbol{\theta}}_{-i}$, is $tr_j\left(\widetilde{K}_j \big| K_j, \boldsymbol{\eta}_i, \widehat{\boldsymbol{\eta}}_{-i}\right)$. Hence, we can now use the formula for the payments to obtain:

$$
Er_i\left(\widehat{\boldsymbol{\theta}}; \boldsymbol{\theta}\right) = \sum_{j \in \mathcal{I} \backslash i}\left[\begin{array}{c} \sum_{\widetilde{K}_j \subseteq K_j^*\left(\widehat{\boldsymbol{\theta}}_i, \widehat{\boldsymbol{\theta}}_{-i}\right)} \widehat{v}_j\left(\widetilde{K}_j\right) \cdot tr_j\left(\widetilde{K}_j \big| K_j^*\left(\widehat{\boldsymbol{\theta}}_i, \widehat{\boldsymbol{\theta}}_{-i}\right), \boldsymbol{\eta}_i, \widehat{\boldsymbol{\eta}}_{-i}\right) \\ - \widehat{c}_j\left(K^*\left(\widehat{\boldsymbol{\theta}}_i, \widehat{\boldsymbol{\theta}}_{-i}\right)\right) \end{array}\right] - B_i(\widehat{\boldsymbol{\theta}}_{-i})
\tag{19}
$$

If we replace the expression above into the formula for $\overline{u}_i$ we can observe that an agent can only affect its utility with its report by changing $K^*(\widehat{\boldsymbol{\theta}})$. The key point to note is that the agent computes the value of the trust function using the true value of $\boldsymbol{\eta}_i$ (rather than its reported value $\widehat{\boldsymbol{\eta}}_i$).

Now, Equation (16) implies that for all allocations $K$:

$$
\overline{u}_i\left(\boldsymbol{\theta}_i, \widehat{\boldsymbol{\theta}}_{-i}; \boldsymbol{\theta}\right) \geq \overline{u}_i\left(\widehat{\boldsymbol{\theta}}_i, \widehat{\boldsymbol{\theta}}_{-i}; \boldsymbol{\theta}\right),
\tag{20}
$$

because the efficient allocation, computed by taking into account $i$'s true type $\boldsymbol{\theta}_i$ and the reported types of all other agents $\widehat{\boldsymbol{\theta}}_{-i}$ is better than or equal to any other allocation.

Given the above condition and since Equation (20) applies to all possible realisations of $\boldsymbol{\theta}$, the mechanism is incentive compatible. $\square$

**Proposition 5.** *The GTBM is efficient.*

*Proof.* Given the incentive compatibility of the mechanism, the centre will receive truthful reports from all the agents. As a result, it will compute the allocation according to Equation (16), thereby leading to an efficient outcome. $\square$

**Proposition 6.** *There exist values of $B_i(\cdot)$ such that the GTBM is individually rational.*

*Proof.* We again begin by making the standard assumption that the agent derives $u_i = 0$, when not participating in the mechanism. Then, it remains to be shown that the agent derives non-negative utility from the mechanism. Since the efficient allocation is chosen (and is at worst a null allocation), the expected utility of each agent is always greater than or equal to $-B_i(\cdot)$ according to Equation (18). Since $B_i(\cdot)$ can be set to 0, the mechanism is individually rational. $\square$

Note that there are possibly many other values of $B_i(\boldsymbol{\theta}_{-i})$, besides $B_i = 0$, that guarantee individual rationality.

Speaking more generally, it can easily be seen that the GTBM mechanism of the multiple task, multiple requester scenario is a generalisation of the GTBM mechanism with a single requester and a single task. It is also a generalisation of the mechanism of Porter et al. where they simply assume that each agent only has an EQOS about its own probability of success. Moreover, in the paper by Porter et al. for example, $B_i$ is specified as follows:

$$B_i(\widehat{\boldsymbol{\theta}}_{-i}) = \max_{K' \in \mathcal{K}_{-i}} \left( v_0(K') \cdot \widehat{p}_{-i}(\tau) - \sum_{j \in \mathcal{I} \setminus i} \widehat{c}_j(K') \right)$$

where $\widehat{p}_{-i}(\tau)$ is the reported probability of completion of the agent assigned the task in allocation $K'$ and $\mathcal{K}_{-i}$ is the set of allocations excluding agent $i$.

### 4.4 Extracting the Minimum Marginal Contribution

Up to now, we have considered that $B_i(\boldsymbol{\theta}_{-i})$ could be set to arbitrary values to try and reduce the payments made by the centre to all the agents. More interestingly, it should be possible, as in the standard VCG mechanism, to only pay an agent its marginal contribution to the system. However, in our case, due to the interdependence of valuations, it is not as simple as comparing the social welfare with and without a given agent in the system as is commonly done in VCG-based mechanisms (Porter et al., 2008 is an obvious example of this). This is because, in our case, when an agent is removed from the domain used to compute the efficient allocation, the remaining EQOS reports can arbitrarily change the allocation value. This could, in turn, be exploited by other agents to improve their utility. The example in Section 3.3.1 showing the failure of a simple extension of Porter et al.'s mechanism illustrates this point.

Assuming that the centre wants to induce participation by all agent types, here we propose a novel approach to extracting the marginal contribution of an agent, by taking into account EQOS reports of other agents and *possible reports* that the agent could make. Let $\mathcal{K}_{-i}$ be the set of possible allocations when agent $i$ is excluded from society. The value of $B_i(\cdot)$ can be chosen such that it is equivalent to the social utility of the mechanism when agent $i$ is excluded and its EQOS reports are chosen so as to minimise social utility, that is:

$$B_i(\widehat{\boldsymbol{\theta}}_{-i}) = \min_{\boldsymbol{\eta}_i \in [0,1]^{|\mathcal{I}| \times |\mathcal{T}|}} \max_{K \in \mathcal{K}_{-i}} \sum_{j \in \mathcal{I} \setminus i} \left[ \sum_{\widetilde{K}_j \subseteq K_j} \widehat{v}_j \left( \widetilde{K}_j \right) \cdot tr_j \left( \widetilde{K}_j \,\middle|\, K_j, \boldsymbol{\eta}_i, \widehat{\boldsymbol{\eta}}_{-i} \right) - \widehat{c}_j \left( K \right) \right] \quad (21)$$

It is to be noted that $B_i$ is computed using the lowest trust values that could be derived using $i$'s EQOS reports.

Then, the generalised payment scheme is:

$$\begin{aligned}
r_i \left( \widehat{\boldsymbol{\theta}}, \kappa(.) \right) = \sum_{j \in \mathcal{I} \setminus i} & \left[ \sum_{\widetilde{K}_j \subseteq K_j^* \left( \widehat{\boldsymbol{\theta}} \right)} \widehat{v}_j \left( \widetilde{K}_j \right) \cdot \kappa \left( \widetilde{K}_j \right) - \widehat{c}_j \left( K^* \left( \widehat{\boldsymbol{\theta}} \right) \right) \right] \\
& - \min_{\boldsymbol{\eta}_i \in [0,1]^{|\mathcal{I}| \times |\mathcal{T}|}} \max_{K \in \mathcal{K}_{-i}} \sum_{j \in \mathcal{I} \setminus i} \left[ \sum_{\widetilde{K}_j \subseteq K_j} \widehat{v}_j \left( \widetilde{K}_j \right) \cdot tr_j \left( \widetilde{K}_j \,\middle|\, K_j, \boldsymbol{\eta}_i, \widehat{\boldsymbol{\eta}}_{-i} \right) - \widehat{c}_j \left( K \right) \right]
\end{aligned}$$
$$(22)$$

The point to note here is that incentive compatibility (and hence efficiency of the mechanism) still holds given that the payment scheme is still independent of $i$'s reports. In fact, $r_i$ rewards $i$ with the maximum difference that agent $i$ could make by setting all elements in $\boldsymbol{\eta}_i$ to different values in $[0,1]^{|\mathcal{I}|\times|\mathcal{T}|}$.[8]

This procedure reduces the payments made by the centre, while keeping individual rationality since the value of the efficient allocation (given incentive compatibility as proven earlier) is always higher than or equal to the value of $B_i$, which means that:

$$\overline{u}_i\left(K^*(\boldsymbol{\theta}),\boldsymbol{\theta}\right) = \sum_{j\in\mathcal{I}}\left[\sum_{\widetilde{K}_j\subseteq K_j^*(\boldsymbol{\theta})} v_j\left(\widetilde{K}_j\right)\cdot tr_j\left(\widetilde{K}_j\,\middle|\,K_j^*(\boldsymbol{\theta}),\boldsymbol{\eta}\right) - c_j\left(K^*(\boldsymbol{\theta})\right)\right]$$

$$- \min_{\boldsymbol{\eta}_i\in[0,1]^{|\mathcal{I}|\times|\mathcal{T}|}}\max_{K\in\mathcal{K}_{-i}}\sum_{j\in\mathcal{I}\setminus i}\left[\sum_{\widetilde{K}_j\subseteq K_j} v_j\left(\widetilde{K}_j\right)\cdot tr_j\left(\widetilde{K}_j\,\middle|\,K_j,\boldsymbol{\eta}_i,\boldsymbol{\eta}_{-i}\right) - c_j\left(K\right)\right] \geq 0;$$

It is also to be noted that the above equation implies that there is no restriction placed on the functional form of the trust function $tr$ for the payment scheme to work and for the properties of the mechanism to hold. This is an improvement on previous mechanisms (see Section 2) which had considered trust functions that are only monotonically increasing in $\boldsymbol{\eta}_i$ for each $i$.

Now, the choice of $B_i$ determines whether the centre runs the mechanism at a profit or not. Hence, to understand what the scale of payments may be in the GTBM discussed in this section, consider the following example.

**Example 8.** *There are $n$ agents, $\mathcal{I} = \{1,...,n\}$, each requiring that a single task be performed for them. All agents have value 1 for the task to be performed for them and have zero cost for performing all tasks. The EQOS of agent $h$ about agent $i$'s probability of succeeding at the task for agent $j$ is $\eta_h^i(K_j)\in[x,1]$ for all $h,i,j = 1,...,n$. Suppose that $tr_j^i(\boldsymbol{\eta}) = \left[\sum_{h\in\mathcal{I}}\eta_h^i(K_j)\right]/n$.*

In the above example, the EQOS of each agent are in the interval $[x,1]$, so that $x$ can be viewed as the lower bound on the expected probability of success at each task. From Equation (21) we can compute the value of $B_i$:

$$B_i(\boldsymbol{\theta}_{-i}) = \sum_{j\in\mathcal{I}\setminus i}\max_{\ell\in\mathcal{I}}\left[\frac{\sum_{h\in\mathcal{I}\setminus i}\eta_h^\ell(K_j)+x}{n}\right]$$

Note that, depending on the value of $\boldsymbol{\eta}_{-i}$, $B_i(\boldsymbol{\theta}_{-i})$ could be any value between $(n-1)x$ and $(n-1)(n-1+x)/n$; $B_i$ increases with the lower bound $x$ on agents' EQOS. The actual payment to agent $i$ will depend on the success or failure of each task (e.g., the payment is $-B_i$ if all tasks fail). From Equation (19), we can calculate the value of the expected payment to agent $i$ as:

$$Er_i(\boldsymbol{\theta}) = \sum_{j\in\mathcal{I}\setminus i}\max_{\ell\in\mathcal{I}}\left[\sum_{h\in\mathcal{I}}\frac{\eta_h^\ell(K_j)}{n}\right] - B_i(\boldsymbol{\theta}_{-i})$$

$$\leq \sum_{j\in\mathcal{I}\setminus i}\left[\frac{\eta_i^{\ell^*(j)}-x}{n}\right]$$

---

8. This minimisation takes place over the domain of trust values which could be other than $[0,1]$ in the general case.

where $\ell^*(j)$ is the agent allocated the task for agent $j$ under the efficient allocation rule. Let $EV$ be the total expected value from all tasks:

$$EV(\boldsymbol{\theta}) = \sum_{j \in \mathcal{I}} \left[ \frac{\sum_{i \in \mathcal{I}} \eta_i^{\ell^*(j)}}{n} \right]$$

Note that the total expected value of all tasks is greater than the sum of the expected payments over all agents, that is:

$$\sum_{j \in \mathcal{I}} \left[ \frac{\sum_{i \in \mathcal{I}} \eta_i^{\ell^*(j)}}{n} \right] > \sum_{i \in I} \left[ \sum_{j \in \mathcal{I} \setminus i} \left[ \frac{\eta_i^{\ell^*(j)} - x}{n} \right] \right]$$

Thus, the centre always profits from the mechanism. A lower bound on the difference between total expected value and total expected payments is $[\sum_{i \in \mathcal{I}} \eta_i^{\ell^*(j)} + (n-1)x]/n$. Note also that the lower bound on the centre's profit from the mechanism increases with the lower bound on EQOS $x$.

As we pointed out in the discussion of Example 6 in Section 3.3.2, if the centre is trying to minimise payments, it could give up on individual rationality, by increasing $B_i$, at the cost of inducing some agent types not to participate in the mechanism. This may be appealing when the probability of task failure is high; in such cases, the centre may prefer to avoid paying an amount almost as large as the total value of the tasks. On the other hand, in a number of practical applications the centre may want to use the mechanism that induces participation by all types, described in this section. This is certainly the case, for example, if the lower bound on EQOS (i.e., the lower bound on the probability that tasks are successful) is high. Moreover, our mechanism with participation by all types is appropriate when the centre mainly seeks to maximise social welfare. Consider, for example, a government that is trying to boost the economy through major public infrastructure projects. In order to do so, it may be willing to invest in the trust-based mechanism to get the best infrastructures built at the cheapest cost. Moreover, the government may be willing to make a low profit in order to ensure the survivability of the construction companies by guaranteeing them some payoff if they participate in the mechanism. Another example where a company might want to involve all task performers would be a company trying to acquire as much information as possible about all task performers in order to maximise the returns on its future decisions. Following from our running scenario, say MoviePictures.com needs to contract a video editing company to add computer graphics to a movie that may become a blockbuster if the graphics are well done. In case the task is successful, MoviePictures.com is likely to get many contracts in the future. It is therefore critical that all the available information is collected from agents in order to choose the most reliable video editing company. In this case, MoviePictures.com may accept a smaller short-run profit by running our mechanism with full participation, in order to guarantee that the selected agent is the best one and that future contracts will be obtained.

To summarise, in this section we have devised a mechanism that is incentive compatible, individually rational and efficient for task allocation under uncertainty when multiple distributed reports are used in order to judge this uncertainty. It is to be noted that we did not need two-stage mechanisms, as in the work of Mezzetti (2004), because in our settings we can condition payments on the completion of the tasks (the indicator function $\kappa(\cdot)$ captures this dependence of payments on task completion). So far, we have just considered the economic properties of the mechanisms, but as we argued earlier, this is only part of the picture. In the next section, we report on its implementation.

## 5. Implementing the Generalised Trust-Based Mechanism

As shown above, the addition of trust to the basic task allocation problem not only complicates the payment scheme, but also requires a larger number of important optimisation steps than the normal VCG. In more detail, trust-based mechanisms require that agents specify an expected value for a set of tasks depending on the performer of such tasks which, in turn, means that the space of solutions to be explored is significantly larger than in common task allocation problems. Moreover, the payment scheme of trust-based mechanisms requires finding the efficient allocation multiple times with and without the agents' reports. With this added level of complexity, it is important to show that the mechanisms are actually implementable and that solutions can be found for usefully sized problems in reasonable time.[9]

Against the above background, in this section we describe the first formulation and implementation of the GTBM. In particular, in the GTBM, we tackle the main optimisation problem posed by Equation (16) (which is then repeated several times in the payment scheme). This is commonly referred to as the winner determination problem in combinatorial auctions. In order to solve it, we take insight from solutions to combinatorial exchanges which often map the problem to a well studied matching problem (Kalagnanam & Parkes, 2004; Engel, Wellman, & Lochner, 2006). In so doing, we develop a novel representation of the optimisation problem by using hypergraphs to describe the relationships between valuations, trust, and bids by task performers and then cast the problem as a special hypergraph matching problem. Given this representation, we are then able to solve the problem using Integer Programming techniques through a concise formulation of the objective function and constraints.

### 5.1 Representing the Search Space

It is important to define the search space in such a way that relationships between valuations, bids, trust, and tasks can be clearly and concisely captured. In particular, our representation aims to map the GTBM optimisation problem to a matching problem that has been well studied in the literature. To do this, the representation must allow us to define the whole space of feasible task allocations, and, subsequently, define how to select them as valid solutions to the GTBM optimisation problem. Now, to allow bidders (task performers) and askers (task requesters) to express their bids and valuations in a consistent and implementable way, we choose the XOR bidding language. Such a bidding language requires that an auctioneer can accept at most one bid out of each XOR bid and that each XOR bid can belong to only one agent. We choose this particular bidding language because it has been shown that any valuation can be expressed using it (Nisan, 2006).[10] An example of an XOR bid in our context would be $\{c_i(\tau_1, \tau_2) \; XOR \; c_i(\tau_1, \tau_3) \; XOR \; c_i(\tau_1, \tau_2, \tau_3)\}$ which means that agent $i$ would only go for one of these three bids over tasks $\tau_1, \tau_2$ and $\tau_3$ ($c_i$ could also be replaced by $v_i$ for task requesters). In terms of our running example, such a bid would express PoorRender Ltd's cost for performing a sound editing task (i.e., $\tau_1$), a movie production task (i.e., $\tau_2$), or both in combination (i.e., $\tau_1, \tau_2$).

---

9. It is already known that computing the efficient allocation and payments for VCG mechanisms is NP-hard (Sandholm, Suri, Gilpin, & Levine, 2002). Therefore, finding efficient solutions to VCG mechanisms is already a significant challenge in its own right.

10. Other bidding languages (such as those describing Atomic or OR bids, as in Nisan, 2006) could equally well be used in our model and would only require minor changes to the constraints that we need to apply.

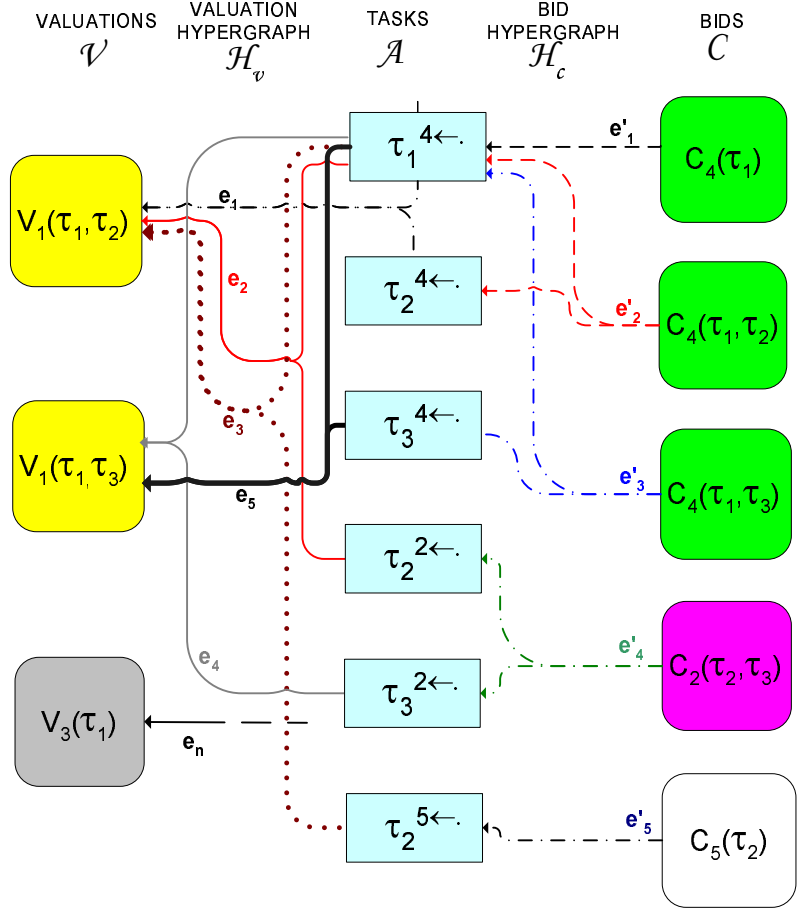

*Figure 1:* Graphical representation of the GTBM search space. Nodes of the same colour represent valuation or cost nodes that belong to the same agent (here nodes with $v_1$ belong to agent 1 and those with $c_4$ belong to agent 4). Edges of the same colour either originate from the same node or end up at the same node.

To build the overall representation of the problem, we first focus on representing expected valuations and costs as well as their relationships. These are depicted in Figure 1. In more detail, we specify three types of nodes: (1) valuations (along the $\mathcal{V}$ column); (2) bids (under the $\mathcal{C}$ column); and (3) task-per-bidder nodes (under the $\mathcal{A}$ column). Each node $v_i(\boldsymbol{\tau})$ in the $\mathcal{V}$ column stands for a valuation submitted by agent $i$ over a set of tasks $\boldsymbol{\tau} \subseteq \mathcal{T}$. Each node $c_j(\boldsymbol{\tau})$ in the $\mathcal{C}$ column stands for a bid issued by agent $j$ over tasks $\boldsymbol{\tau} \subseteq \mathcal{T}$. Each element of $\mathcal{A}$ represents the allocation $\tau_m^{j \leftarrow \cdot}$ of a single task $\tau_m \in \mathcal{T}$ to task performer (bidder) $j$ by a task requester yet to be determined (represented by a dot). In other words, the elements in $\mathcal{A}$ represent *patterns* for single-task allocations. We term such elements task-per-bidder nodes.

Note that it is possible that different valuations come from the same requester. If so they are labelled by the same subscript. Moreover, since we have opted for an XOR bidding language, valuations belonging to the very same requester are mutually exclusive.

### 5.1.1 Defining Relationships between Valuations, Tasks, and Bids

Given the nodes defined by $\mathcal{A}$, $\mathcal{V}$, and $\mathcal{C}$, by relating a node $\tau_m^{j\leftarrow\cdot}$ in $\mathcal{A}$ to a node $v_i(...,\tau_m,...)$ in $\mathcal{V}$ we define the assignment of task $\tau_m$ by $i$ to $j$ through the specific valuation $v_i(...,\tau_m,...)$. Similarly, by relating a node $\tau_m^{j\leftarrow\cdot}$ in $\mathcal{A}$ to a node $c_j(...,\tau_m,...)$ in $\mathcal{C}$ we define the assignment of the task to the specific bid $c_j(...,\tau_m,...)$ by agent $j$. Therefore, a triple $(v,\tau_m^{j\leftarrow\cdot},c)$ where $v\in\mathcal{V},\tau_m^{j\leftarrow\cdot}\in\mathcal{A},c\in\mathcal{C}$ fully characterises an allocation for task $\tau_m$, namely a *single-task allocation*. Hence, as can be seen in Figure 1, we define two types of relationships: between valuations and task-per-bidder nodes (noted by edges $e_1,e_2,...$), and between bids and task-per-bidder nodes (noted by edges $e_1',e_2',...$).[11]

Using these relationships, a valuation can then be related to a *set of task-per-bidder nodes* if and only if these fully cover the performance of the task(s) in the valuation. For instance, we can relate $v_1(\tau_1,\tau_2)$ to nodes $\tau_1^{4\leftarrow\cdot}$ (agent 4 performs task $\tau_1$) and $\tau_2^{2\leftarrow\cdot}$ (agent 2 performs task $\tau_2$) because they guarantee the performance of tasks $\tau_1$ and $\tau_2$. Similar to valuation relationships, each node $\mathcal{C}$ is only related to the set of task-per-bidder nodes in $\mathcal{A}$ into which each bid splits. Thus, in Figure 1, bid $c_4(\tau_1)$ is only related to $\tau_1^{4\leftarrow\cdot}$, whereas bid $c_2(\tau_2,\tau_3)$ is related to nodes $\tau_2^{2\leftarrow\cdot}$ and $\tau_3^{2\leftarrow\cdot}$.

Thus, we can identify the task performers for each task in a given valuation. This is critical since the GTBM, contrary to common task allocation mechanisms (such as VCG or $\text{M}^{th}$ price auctions), requires that we identify exactly who performs a task in order to determine the POS of that task (by virtue of the requester's trust in the performer) and hence the expected value of the task.

As can be seen, our representation allows us to capture all tasks and performers of such tasks since each valuation node in $\mathcal{V}$ can be potentially related to multiple nodes in $\mathcal{A}$; and, likewise, each bid in the $\mathcal{C}$ column can be potentially related to multiple nodes in $\mathcal{A}$. To capture these related relationships precisely, we define special edges that can connect several nodes (e.g., the ones depicted as $e_1,e_2,\cdots,e_1',e_2'....$ in Figure 1). Such edges are termed hyperedges because they combine a number of singleton edges. Hence, Figure 1 can be best described as a *hypergraph* (Berge, 1973). In order to precisely define the matching problem that the GTBM poses, we elaborate on the formalism of hypergraphs since this will help in concisely expressing the problem later on. More specifically, the formal notion of hypergraphs, as introduced in the paper by Berge (1973), is:

**Definition 9. Hypergraph.** *Let $X=\{x_1,x_2,\ldots,x_n\}$ be a finite set of $n$ elements, and let $\mathcal{E}=\{e_j|j\in J\}$ be a family of subsets of $X$ where $J=\{1,2,...\}$. The family $\mathcal{E}$ is said to be a hypergraph on $X$ if:*

1. *$e_j\neq\emptyset\ (\forall j\in J)$*

2. *$\cup_{j\in J}e_j=X$.*

*The pair $H=(X,\mathcal{E})$ is called a hypergraph. The elements $x_1,x_2,\ldots,x_n$ are called the vertices and the sets $e_1,e_2,\ldots,e_j$ are called the hyperedges.*

We say that a hypergraph is *weighted* if we associate to each hyperedge $e\in\mathcal{E}$ a real number, $w(e)$, called the *weight* of $e$. This is used to give more or less importance to some edges.

From the formal definition of hypergraphs, we observe that Figure 1 results from the overlapping of two separate hypergraphs: (i) the *valuation hypergraph* that occurs from linking valuations to task-per-bidder nodes; and (ii) the *bid hypergraph* that occurs from linking each bid to the corresponding task-per-bidder nodes. In what follows, we formally define both hypergraphs from valuations and bids so that later on we can structurally characterise the notions of feasible and optimal allocations.

---

11. Figure 1 only depicts a sample of all possible relationships for ease of illustration.

### 5.1.2 THE VALUATION HYPERGRAPH

The valuation hypergraph highlights the main difference between the GTBM and the common combinatorial exchanges (e.g., those based on traditional VCG or M$^{th}$-price auctions). In particular, in the GTBM valuations need to take into account the trust of the task requester in the task performer while, in normal combinatorial exchanges, task requesters are indifferent to task performers. This means the weight of each hyperedge in a valuation hypergraph is dependent on trust and a large number of edges need to be generated (one per task performer) which is not the case in normal combinatorial exchanges.

To define the valuation hypergraph, we need to define hyperedges that emanate from each node in $\mathcal{V}$ to one or more nodes in $\mathcal{A}$. To this end, let $\mathcal{V} = \{v_i(\boldsymbol{\tau}) \neq 0 | \boldsymbol{\tau} \subseteq \mathcal{T}, i \in \mathcal{I}\}$ and $\mathcal{C} = \{c_j(\boldsymbol{\tau}) \neq \infty | \boldsymbol{\tau} \subseteq \mathcal{T}, j \in \mathcal{I}\}$ be the sets of all valuations and all bids respectively. Let $\boldsymbol{\tau}^{j\leftarrow\cdot} = \{\tau \in \mathcal{T} \mid \exists \boldsymbol{\tau}' \subseteq \mathcal{T} : c_j(\boldsymbol{\tau}) \neq \infty \text{ and } \tau \in \boldsymbol{\tau}'\}$ be the set of tasks over which agent $j$ submits bids. Hence, $\mathcal{A} = \{\tau_k^{j\leftarrow\cdot} | \tau_k \in \boldsymbol{\tau}^{j\leftarrow\cdot}, j \in \mathcal{I}, c_j(\boldsymbol{\tau}) \in \mathcal{C}\}$ is the set containing all the tasks bid by each bidder.[12]

Furthermore, we need to define some auxiliary sets as follows. Given a valuation over a set of tasks $\boldsymbol{\tau}$, a set of nodes $A \subseteq \mathcal{A}$ fulfils it if and only if:

$$\bigcup_{\tau_k^{j\leftarrow\cdot} \in A} \{\tau_k^{j\leftarrow\cdot}\} = \boldsymbol{\tau} \text{ and } |\boldsymbol{\tau}| = |A|$$

For instance, the set of nodes $A = \{\tau_1^{4\leftarrow\cdot}, \tau_2^{2\leftarrow\cdot}\}$ fulfils any valuation over $\{\tau_1, \tau_2\}$. Hence, the subsets of $\mathcal{A}$ that fulfil a valuation over a set of tasks $\boldsymbol{\tau}$ are expressed using $A^{\boldsymbol{\tau}}$ which is defined as:

$$A^{\boldsymbol{\tau}} = \{A \subseteq \mathcal{A} \mid \bigcup_{\tau_k^{j\leftarrow\cdot} \in A} \{\tau_k^{j\leftarrow\cdot}\} = \boldsymbol{\tau} \text{ and } |\boldsymbol{\tau}| = |A|\}$$

For instance, considering the example in Figure 1,

$$A^{\{\tau_1,\tau_2\}} = \{\{\tau_1^{4\leftarrow\cdot}, \tau_2^{4\leftarrow\cdot}\}, \{\tau_1^{4\leftarrow\cdot}, \tau_2^{2\leftarrow\cdot}\}, \{\tau_1^{4\leftarrow\cdot}, \tau_2^{5\leftarrow\cdot}\}\}$$
$$A^{\{\tau_1,\tau_3\}} = \{\{\tau_1^{4\leftarrow\cdot}, \tau_3^{4\leftarrow\cdot}\}, \{\tau_1^{4\leftarrow\cdot}, \tau_3^{2\leftarrow\cdot}\}\}$$

Given the above definitions, we can now define the set of all hyperedges connected to a valuation $v_i(\boldsymbol{\tau}) \in \mathcal{V}$ as:

$$\mathcal{E}_i^v(\boldsymbol{\tau}) = \cup_{a \in A^{\boldsymbol{\tau}}} \{\{v_i(\boldsymbol{\tau})\} \cup a\}$$

For instance, from Figure 1:

$$\mathcal{E}_1^v(\tau_1, \tau_2) = \{e_1, e_2, e_3\} \text{ and } \mathcal{E}_1^v(\tau_3) = \{e_4, e_5\},$$

where $e_1 = \{v_1(\tau_1, \tau_2), \tau_1^{4\leftarrow\cdot}, \tau_2^{4\leftarrow\cdot}\}$, $e_2 = \{v_1(\tau_1, \tau_2), \tau_1^{4\leftarrow\cdot}, \tau_2^{2\leftarrow\cdot}\}, \ldots$, and so on.

---

12. Recall that since the mechanism has been proven to be incentive-compatible we can use the agents' true valuations and costs instead of their reported counterparts.

The set of all hyperedges containing valuations of the very same agent $i$ is defined as:

$$\mathcal{E}_i^v = \bigcup_{\boldsymbol{\tau} \subseteq \mathcal{T}} \mathcal{E}_i^v(\boldsymbol{\tau})$$

Then, the set of hyperedges connecting nodes in $\mathcal{V}$ to nodes in $\mathcal{A}$ is defined as:

$$\mathcal{E}^v = \bigcup_{i \in \mathcal{I}} \mathcal{E}_i^v$$

Given this, we define the valuation hypergraph as a pair:

$$\mathcal{H}^v = (\mathcal{V} \cup \mathcal{A}, \mathcal{E}^v)$$

Thus, each hyperedge in $\mathcal{H}^v$ consists of a single valuation vertex corresponding to an element in $\mathcal{V}$ along with a complete task allocation for the valued tasks out of the task-per-bidder nodes in $\mathcal{A}$.

The valuation hypergraph $\mathcal{H}^v$ partly defines the space within which a solution needs to be found. However, in order to define the quality of the solution found, it is important to define the weight attached to each hyperedge of the hypergraph $\mathcal{H}^v$. The weight of a hyperedge is actually equal to the expected value of the allocation of the tasks to a set of task performers (bidders). Consider, for instance, valuation $v_1(\tau_1, \tau_2)$. All the possible matchings that fulfil it are represented by all the pairs $(\tau_1^{\cdot \leftarrow 1}, \tau_2^{\cdot \leftarrow 1})$. For example, the hyperedge $e_2$ involving the pairing $(\tau_1^{4 \leftarrow 1}, \tau_2^{2 \leftarrow 1})$ denotes that agent 4 performs task 1 for agent 1 and agent 2 performs task 2 for agent 1. The expected valuation associated to this allocation depends on the POS of agents 4 and 2 when performing $\tau_1$ and $\tau_2$ respectively.

In this case, the expected valuation associated to $e_2$ is assessed as:

$$\begin{aligned}
\overline{v}_1(\tau_1^{4 \leftarrow 1}, \tau_2^{2 \leftarrow 1}) = {} & v_1(\tau_1, \tau_2) \cdot p_4(\tau_1^{4 \leftarrow 1}) \cdot p_2(\tau_2^{2 \leftarrow 1}) + \\
& v_1(\tau_1) \cdot p_4(\tau_1^{4 \leftarrow 1}) \cdot (1 - p_2(\tau_2^{2 \leftarrow 1})) + \\
& v_1(\tau_2) \cdot (1 - p_4(\tau_1^{4 \leftarrow 1})) \cdot p_2(\tau_2^{2 \leftarrow 1})
\end{aligned} \tag{23}$$

where $p$ is a function that returns the POS of the agent that is assigned a given task (computed using confidence, reputation, or trust). Notice that the value $(1 - p_i(\tau_k^{i \leftarrow j}))$ represents the probability of agent $i$ failing to perform task $\tau_k$ for agent j. Since no requests are submitted for $\tau_1$ and $\tau_2$ alone, $v(\tau_1) = v(\tau_2) = 0$. Thus, the expected valuation associated to the particular allocation represented by arc $e_2$ becomes $\overline{v}_1(\tau_1^{4 \leftarrow 1}, \tau_2^{2 \leftarrow 1}) = v_1(\tau_1, \tau_2) \cdot p_4(\tau_1^{4 \leftarrow 1}) \cdot p_2(\tau_2^{2 \leftarrow 1})$. With a similar argument, we obtain $\overline{v}_1(\tau_1^{4 \leftarrow 1}, \tau_2^{5 \leftarrow 1}) = v_1(\tau_1, \tau_2) \cdot p_4(\tau_1^{4 \leftarrow 1}) \cdot p_5(\tau_2^{5 \leftarrow 1}) \neq \overline{v}_1(\tau_1^{4 \leftarrow 1}, \tau_2^{2 \leftarrow 1})$, corresponding to hyperedge $e_3$.

Generalising, given a hyperedge $e \in \mathcal{E}^v$ with valuation $v_i(\boldsymbol{\tau})$, we can readily build an allocation for the tasks in $\boldsymbol{\tau}$ from the elements in $e$ and $v_i(\boldsymbol{\tau})$. If $p$ is a function that returns the POS (be it confidence, reputation, or trust) of a given task performer from each requester's point of view, then we can compute the expected valuation of the allocation defined by hyperedge $e$ as follows:

$$\overline{v}_i(\boldsymbol{\tau}) = \sum_{\boldsymbol{\tau}' \subseteq \boldsymbol{\tau}} \left( v_i(\boldsymbol{\tau}') \prod_{\tau_l^{j \leftarrow \cdot} \in e, \tau_l \in \boldsymbol{\tau}'} p_j(\tau_l^{j \leftarrow i}) \prod_{\tau_l^{j \leftarrow \cdot} \in e, \tau_w \in \boldsymbol{\tau} \setminus \boldsymbol{\tau}'} \left(1 - p_j(\tau_w^{j \leftarrow i})\right) \right) \tag{24}$$

In other words, given a hyperedge $e \in \mathcal{E}^v$, its weight is assessed using Equation (24) which is equivalent to the expected value computed in Equation (16) (i.e., the sum of expected values over all allocations from agent $i$). Now, given that each edge of the valuation hypergraph is assigned a weight, $\mathcal{H}^v$ is termed a weighted hypergraph.

### 5.1.3 The Bid Hypergraph

To define the bid hypergraph we need to determine the hyperedges that connect bids to task-per-bidder nodes. In more detail, given a bid $c_j(\boldsymbol{\tau}) \in \mathcal{C}$, we relate it to the task-per-bidder nodes in $\mathcal{A}$ by constructing hyperedge $\mathcal{E}_j^c(\boldsymbol{\tau}) = \{c_j(\boldsymbol{\tau})\} \cup \{\tau_k^{j\leftarrow}|\tau_k \in \boldsymbol{\tau}\}$. This hyperedge is assigned a weight which is equal to the cost of $c_j(\boldsymbol{\tau})$. Then the set of all hyperedges containing all the bids of agent $i$ can be defined as:

$$\mathcal{E}_i^c = \bigcup_{\boldsymbol{\tau} \subseteq \mathcal{T}} \mathcal{E}_c^i(\boldsymbol{\tau})$$

Given this, the set of all hyperedges connecting nodes in $\mathcal{C}$ to nodes in $\mathcal{A}$ can be defined as:

$$\mathcal{E}^c = \bigcup_{i \in \mathcal{I}} \mathcal{E}_i^c$$

Finally, we define the bid hypergraph as a pair:

$$\mathcal{H}^c = (\mathcal{A} \cup \mathcal{C}, \mathcal{E}^c)$$

In other words, each hyperedge in $\mathcal{H}^c$ consists of a single bid vertex corresponding to an element in $\mathcal{C}$ along with the corresponding task-per-bidder nodes in $\mathcal{A}$. Notice that our definitions of valuation and bid hypergraphs ensure that each hyperedge in $H^v$ contains a single valuation from $\mathcal{V}$ and each hyperedge in $H^c$ contains a single bid from $\mathcal{C}$.

### 5.1.4 Defining the Matching Problem for the GTBM

Having defined the valuation and bid hypergraphs, we can now structurally characterise the notions of feasible and optimal allocations (these are needed to determine the computational complexity of the problem and define the objective function in particular). For this purpose, we must firstly recall some notions of hypergraph theory. In a hypergraph, two hyperedges are said to be *adjacent* if their intersection is not empty. Otherwise they are said to be *disjoint*. For a hypergraph $H = (X, \mathcal{E})$, a family $\mathcal{E}' \subseteq \mathcal{E}$ is defined to be a *matching* if the hyperedges of $\mathcal{E}'$ are pairwise disjoint. With respect to a given matching $\mathcal{E}'$, a vertex $x_i$ is said to be *matched* or *covered* if there is a hyperedge in $\mathcal{E}'$ incident to $x_i$. If a vertex is not matched, it is said to be *unmatched* or *exposed*. A matching that leaves no vertices exposed is said to be *complete*.

Based on the definitions above, we can characterise feasible allocations in the GTBM as follows. First, we must find a matching for the valuation hypergraph that is not necessarily complete (some valuations may remain exposed). Second, we must find another matching for the bid hypergraph that is not necessarily complete either. The two matchings must be related in the following manner: the task-per-bidder nodes in both matchings should be the same. In other words, given a task-per-bidder node, it must be related to some valuation node and to some bid node, or else be excluded from both matchings. In this way, valuations and bids are linked to create single-task allocations. For instance, in Figure 1, if $e_2$ belongs to the matching for the valuation hypergraph, then $e_4'$ must be part of the matching for the bid hypergraph to ensure that there is a bid for $\tau_2^{2\leftarrow\cdot}$ and that either $e_1'$, $e_2'$, or $e_3'$ are part of the matching for the bid hypergraph to ensure that there is a bid for $\tau_1^{4\leftarrow\cdot}$. More formally:

**Definition 10.** *Feasible allocation. We say that a pair $(\mathcal{E}^{v'}, \mathcal{E}^{c'})$ defines a feasible allocation iff:*

- $\mathcal{E}^{v'}$ *is a matching for $\mathcal{H}^v$.*

- $\mathcal{E}^{c'}$ *is a matching for $\mathcal{H}^c$.*

- $\forall \tau \in \mathcal{A}$: ($\tau$ *is matched by $\mathcal{E}^{v'}$*) $\iff$ ($\tau$ *is matched by $\mathcal{E}^{c'}$*).

Given a feasible allocation $(\mathcal{E}^{v'}, \mathcal{E}^{c'})$ as defined above, it is straightforward to assess the expected utility of all agents within the system as follows:

$$\sum_{e \in \mathcal{E}^{v'}} w(e) - \sum_{e' \in \mathcal{E}^{c'}} w(e')$$

since the weights of the hyperedges in the valuation hypergraph stand for expected valuations and the weights of the hyperedges in the bid hypergraph stand for costs. Solving Equation (16) in the GTBM amounts to finding the feasible allocation that maximises the expected utility of all agents within the system. Therefore, the following definition naturally follows.

**Definition 11.** *GTBM Task Allocation Problem The problem of assessing the task allocation that maximises the expected utility of all agents within the system amounts to solving:*

$$\underset{(\mathcal{E}^{v'}, \mathcal{E}^{c'})}{\arg\max} \sum_{e \in \mathcal{E}^{v'}} w_v(e) - \sum_{e' \in \mathcal{E}^{c'}} w_c(e') \tag{25}$$

*where $(\mathcal{E}^{v'}, \mathcal{E}^{c'})$ stands for a feasible allocation.*

Having defined the matching problem for the GTBM, we next describe our solution to this problem using Integer Programming techniques that are commonly used to solve such problems (Cerquides, Endriss, Giovannucci, & Rodríguez-Aguilar, 2007).[13]

## 5.2 An Integer Programming Solution

In this section we show how to map the problem posed by Equation (25) into an integer program (Papadimitriou & Steiglitz, 1982) so that it can be efficiently implemented and solved. Given this translation, the resulting program can be solved by powerful commercial solvers such as ILOG CPLEX[14] or LINGO.[15]

---

13. Other special purpose algorithms (e.g., using dynamic programming or search trees) could also be designed to solve this combinatorial problem. However, to understand the magnitude of the problem and to compare the difficulty of solving this problem against other similar problems, we believe it is better to first attempt to find the solution using standard techniques such as IP.

14. http://www.ilog.com

15. http://www.lindo.com

### 5.2.1 OBJECTIVE FUNCTION AND SIDE CONSTRAINTS

The translation of Equation (25) into an IP is reasonably straightforward given our representation. Thus, solving the GTBM task allocation problem amounts to maximising the following objective function:

$$\sum_{e \in \mathcal{E}^v} x_e \cdot w_v(e) - \sum_{e' \in \mathcal{E}^c} y_{e'} \cdot w_c(e') \qquad (26)$$

where $x_e \in \{0, 1\}$ is a binary decision variable representing whether the valuation in hyperedge $e$ is selected or not, and $y_{e'} \in \{0, 1\}$ is a binary decision variable representing whether the bid in hyperedge $e'$ is selected or not. Thus, $x_e$ is a decision variable that selects a given valuation with a given task-bidder matching, and $y_{e'}$ selects a given bid.

However, some side constraints must be fulfilled in order to obtain a valid solution. First, the semantics of the bidding language must be satisfied. Second, if a hyperedge containing a set of task-per-bidder nodes in $\mathcal{A}$ is selected, we must ensure that the bids covering such nodes are selected too. Moreover, as we employ the XOR bidding language, the auctioneer — the centre in our case — can only select at most one bid per bidder and at most one valuation per asker. Thus, as for bidders, this constraint translates into:

$$\sum_{e' \in \mathcal{E}^c_i} y_{e'} \leq 1 \quad \forall i \in \mathcal{I} \qquad (27)$$

For instance, in Figure 1 this constraint ensures the auctioneer selects one hyperedge out of $e'_1, e'_2$, and $e'_3$, since they all belong to agent 4 (they all come from nodes labelled with the same subscript $c_4(.)$).

For the valuations, the XOR constraints involving them are collected in the following expression:

$$\sum_{e \in \mathcal{E}^v_i} x_e \leq 1 \quad \forall i \in \mathcal{I} \qquad (28)$$

For instance, in Figure 1 this constraint forces the auctioneer to select one hyperedge out of $e_1, e_2$, $e_3, e_4$, and $e_5$ since they all belong to agent 1 (they all come from nodes labelled with the same subscript $v_1(.)$).

If a valuation hyperedge $e \in \mathcal{E}^v$ is selected, the set of task-per-bidder nodes in $\mathcal{A}$ connected to $e$ must be performed by the corresponding bidder agent. For instance, in Figure 1, if hyperedge $e_5$ is selected, the task-per-bidder nodes $\tau_1^{4 \leftarrow 1}$ and $\tau_3^{4 \leftarrow 1}$ must be covered by some bid of agent 4. In this case, bid $c_4(\tau_1, \tau_3)$ is the one covering those tasks. Thus, if we select hyperedge $e_5$ we are forced to select bid $c_4(\tau_1, \tau_3)$ by selecting hyperedge $e'_3$. Thus, in terms of hyperedges, we must ensure that the number of valuation hyperarcs containing a given task-per-bidder node is less than or equal to the number of bid hyperarcs containing it. Graphically, this means that the number of incident valuation hyperedges in a given node $a \in \mathcal{A}$ must be less than the number of incident bid hyperedges in $a$.

$$\sum_{e \in \mathcal{E}^v, \tau_k^{j \leftarrow \cdot} \in e} x_e \leq \sum_{e' \in \mathcal{E}^c, \tau_k^{j \leftarrow \cdot} \in e'} y_{e'} \qquad \forall \tau_k^{j \leftarrow \cdot} \in \mathcal{A} \qquad (29)$$

In case of no free-disposal (i.e., if we do not allow agents to execute tasks without them being asked for) we simply have to replace $\leq$ with $=$. To summarise, solving the GTBM task allocation problem amounts to maximising the objective function defined by expression (26) subject to the constraints in expressions (27), (28), and (29). Next, we determine the complexity results for this problem.

### 5.2.2 COMPLEXITY RESULTS

Having represented the GTBM task allocation problem and defined the corresponding IP formulation, we analyse its computational complexity in order to show the difficulty in solving the GTBM. We also identify the main parameters that affect the computational costs of finding the optimal allocation. These parameters should then allow us to determine in which settings the GTBM can be practically used.

**Proposition 7.** *The GTBM task allocation problem is $\mathcal{NP}$-complete and cannot be approximated to a ratio $n^{1-\epsilon}$ in polynomial time unless $\mathcal{P} = \mathcal{ZPP}$, where $n$ is the total number of bids and valuations.*

*Proof.* Notice that our optimisation model, as formalised by Equation (26), naturally translates to a combinatorial exchange (Kalagnanam, Davenport, & Lee, 2000). This translation can be achieved using our representation by taking the goods (in a combinatorial exchange) to be the dummy tasks $\tau \in \mathcal{T}$, the bids the elements in $\mathcal{C}$, and the asks the weights of the hyperedges in $\mathcal{H}_v$. Thus, while bids remain the same in the exchange, the number of valuations may significantly increase. The reason being that the introduction of trust in our theoretical model makes the initial valuations (asks), the elements in $\mathcal{V}$, allocation-dependent. Hence, every single valuation in $\mathcal{V}$ causes several asks to be originated for the exchange when considering the bidder to which each task may be allocated (see examples in Section 5.1.2). As shown by Sandholm et al. (2002), the decision problem for a binary single-unit combinatorial exchange winner determination problem is $\mathcal{NP}$-complete and the optimisation problem cannot be approximated to a ratio $n^{1-\epsilon}$ in polynomial time unless $\mathcal{P} = \mathcal{ZPP}$, where $n$ is the number of bids. Therefore, the optimisation problem is $\mathcal{NP}$-hard, and so it is in GTBM. □

From the above proof, it can be understood that the search space in the GTBM task allocation problem is significantly larger than in traditional combinatorial exchanges because of the dependency of valuations on the bidders performing tasks. In what follows, we provide a formula that allows us to calculate exactly how big this search space is. This allows us to determine whether the instance to be solved can actually be handled by the solver (which will have its own limits on memory requirements and computation time).

In more detail, say that $\mathcal{A}_k$ is the subset of $\mathcal{A}$ containing the task-per-bidder nodes referring to the same tasks. More formally, $\mathcal{A}_k = \{\tau_k^{j \leftarrow} \in \mathcal{A} \mid j \in \mathcal{I}\}$. From the example in Figure 1, $\mathcal{A}_2 = \{\tau_2^{4 \leftarrow}, \tau_2^{2 \leftarrow}, \tau_2^{5 \leftarrow}\}$. Thus, the expression to assess the number of feasible allocations is:

$$|\mathcal{E}^v| = \sum_{i \in \mathcal{I}} \sum_{v_i(\boldsymbol{\tau}) \neq 0} \prod_{\tau_k \in \boldsymbol{\tau}} |\mathcal{A}_k| \qquad (30)$$

Observe that the number of possible allocations can be computed as the cardinality of $\mathcal{E}^v$ (i.e., the number of valuation hyperarcs) since it exactly determines the number of ways the valuations can be satisfied by the provided bids. The total number of decision variables of the Integer Program is thus $|\mathcal{E}^v| + |\mathcal{E}^c|$. Since the number of expected valuations is several times larger than the number of bids, we expect the number of decision variables associated to bid hyperedges to be much less than the number of valuation hyperedges. Hence, assuming that $|\mathcal{E}^c| \ll |\mathcal{E}^v|$, the number of decision variables will be of the order of $|\mathcal{E}^v|$.

In order to understand the implications of these parameters, consider the case in which all task performers bid over *all* tasks and *all* requesters submit a *single* valuation over all tasks. Specifically, consider a scenario with 15 task performers, 20 requesters, and 5 tasks. Given that in this case $|\mathcal{A}_k = 5|$, the number of allocations is $|\mathcal{E}_v| = 20 \times 15^5 = 15187500$. In reality, agents may not be able to submit bids and asks over all tasks and this would result in a significantly lower number of allocations (given the possible matchings). Hence, to see whether such instances can be practically solved, in Appendix A, we report the running times of the solver, showing that instances with less than $2 \times 10^5$ variables can be comfortably solved within 40 seconds (in the worst case). When taken together, our empirical results and our formula to compute the size of the input (i.e., Equation 30) allow us to affirm that, even if the computational cost associated to the GTBM has the potential to be rather high, our solution can handle small and medium sized problems in reasonable time (see table 3). However, as can be seen, the time to complete grows exponentially with the number of tasks. During our experimental analysis, we also found that the impact of increasing the number of task performers and task requesters was not as significant as increasing the number of tasks. This can be explained by the fact that, given our setup, a larger number of tasks allows significantly more matchings between bids and asks than a larger number of bids and asks. Hence, many more task requesters and performers can be accommodated for small numbers of tasks. It should also be noted that we expect these worst case results to occur fairly rarely on average (much less than half of the instances generated from the same parameters), as shown in Figure 2 in Appendix A.

| Set | Tasks | Task Requesters | Task Performers | Worst Case Running Time |
|-----|-------|-----------------|-----------------|-------------------------|
| 1 | 5 | 20 | 15 | 34 s |
| 2 | 8 | 20 | 15 | 40 mins |
| 3 | 10 | 20 | 15 | 3 days |

*Table 3:* Average running times for different numbers of tasks and agents (taken over 300 sample runs for set 1, 50 sample runs for sets 2 and 3).

Having described the complete picture of the GTBM and its implementation, we next discuss some important issues that may arise when trying to use a GTBM for task allocation.

## 6. Discussion

In this paper we have developed task allocation mechanisms that operate effectively when agents cannot reliably complete tasks assigned to them. Specifically, we have designed a novel Generalised Trust-Based Mechanism that is efficient and individually rational. This mechanism deals with the case where task requesters form their opinions about task performers using reports from their environment and their own direct interactions with the performers. In addition to studying the economic properties of the allocation mechanisms, we provided the optimisation model that generates the solutions that guarantee the efficiency of our mechanism. This optimisation model is the first solver for trust-based mechanisms (and other mechanisms in which the value of an allocation depends on the performer of the allocation) and is based on Integer Programming. As a result, we have shown that the input explodes combinatorially due to the huge number of possible allocations that must be enumerated. Nevertheless, while the computational cost associated to the GTBM is shown to be rather high, given our implementation, we are still able to manage small to medium-sized problems in reasonable time.

Speaking more generally, our work on trust-based mechanisms has a number of broader implications. First, the GTBM shows how to explicitly blend work on trust models with work on mechanism design. Since the mechanism guarantees that certain properties hold for task allocation problems, it can be used as a new, well-founded testbed within which trust models can be evaluated. Up to now, trust models have mainly been tested with randomly generated scenarios and interactions that obey somewhat ad hoc market rules such as those used in the ART testbed (Fullam, Klos, Muller, Sabater, Topol, Barber, Rosenschein, & Vercouter, 2005). Second, our work is the first single-stage interdependent valuations mechanism that is efficient and individually rational (as opposed to Mezzetti's two-stage mechanism). This has been made achievable in the settings we consider by capturing the interdependence between types through the trust function and making the payments to the agents contingent on the *actual execution* of tasks. Another novelty of our approach is that we are able to extract the (maximum) marginal contribution of an agent despite the valuations being interdependent (as we have shown in Section 4.4). Third, our implementation of GTBM highlights the importance of considering the computational aspects of any new mechanism, since these determine whether the mechanism is implementable for realistic scenarios and can indeed bring about its claimed benefits. Our work is a strong statement in this direction since we provide the complete picture of the problem, starting from its representation, through its implementation and sample results, to its complexity analysis.

In practical terms, the GTBM is a step towards building robust multi-agent systems for uncertain environments. In such environments, it is important to aggregate the agents' preferences, while taking into account the uncertainty in order to ensure that the solutions chosen result in the best possible outcome for the whole system. Prior to the GTBM, it was not possible to come up with an efficient solution that would maximise this expected utility. Moreover, the fact that agents can express their perception of the task performers' POS is a new way of building more expressive interactions between buyers and sellers of services (Sandholm, 2007). We believe that the more such perceptions are expressed, the better is the ensuing matching between buyers and sellers and our results are proof of the gain in efficiency this better matching brings about (see sections 3.2.1, 3.3.1, and 4.3).

By introducing GTBM as a new class of mechanisms, this work lays the foundations for several areas of inquiry. To this end, we outline some of the main areas below.

- **Budget Balance**: An important economic property of mechanisms in some contexts is budget balance.[16] However, as mentioned in Section 3.3.2, we have designed our TBMs without considering budget balance. In fact, the GTBM is not budget balanced similar to the VCG and Porter et al.'s mechanism. Now, one possible way of overcoming this problem is to sacrifice either efficiency or individual rationality. In fact, the dAGVA mechanism is a counterpart of the VCG which does indeed sacrifice individual rationality for budget balance (see Section 2). Moreover, Parkes, Kalagnanam, and Eso (2001) develop mechanisms where a number of budget balancing schemes are proposed and near-incentive compatibility is attained by making the payments by the agents as close as possible to those of the VCG ones. Their most effective scheme, the Threshold rule, results in a low loss of incentive-compatibility and it has a relatively high efficiency (around $80\%$). Such budget balance may be useful in situations where the centre cannot run the risk of incurring a loss in generating the efficient outcome for the set of agents in the system. For example, MoviePictures.com may not find it worth injecting money into the system to find the efficient outcome if all its subunits are all nearly equally competitive (both in price and POS). Instead MoviePictures.com might prefer a mechanism that generates a near-efficient outcome by increasing $B_i$ as discussed in Section 4.4. By doing this, the set of agents that participate might be reduced because it is not individually rational for all of them to participate in the mechanism, but, nevertheless, MoviePictures.com may obtain a better outcome. In the future, we will study such trade-offs between the efficiency achieved in the system against the profit made by the centre.

---

16. If a mechanism is budget balanced, it computes transfers in each allocation such that the overall transfer in the system is zero (MasColell et al., 1995). Thus, in a budget balanced mechanism, for each allocation $K$ and associated transfer vector $\boldsymbol{r}$, we have $\sum_{r_i \in \boldsymbol{r}} r_i = 0$.

- **Trust in Task Requesters**: One other potential criticism of mechanisms such as ours is that the task requesters (and the centre) must be trusted to reveal the observed execution of the task (Mezzetti, 2004). However, in our setting, task requesters have a strong incentive to reveal their observations (in case these are not publicly visible) since they would prefer their chosen task performer to be available the next time the mechanism is run. To this end, they must ensure that the task performer does not go bankrupt. As noted in Equations (12) and (17), the task performer would have to pay a significant amount to the centre in case it is reported to fail at its task. Hence, the task requester is better off revealing a successful execution if the task performer is indeed successful.

  Another issue with the trust function used is that weights given to each agent's EQOS report may be uncertain. Thus, in this case, agents may have to learn these weights over multiple interactions. Given this, it is important to develop learning and search techniques that will be able to deal with the large number of possible weights that could be used in these trust functions. These techniques will have to take into account the fact that agents may lose out significantly while exploring the search space.

- **Iterative Mechanisms**: The GTBM is a one-shot mechanism in which the allocation and the payments are calculated given the type of the agents $\{\boldsymbol{v}, \boldsymbol{c}, \boldsymbol{\eta}\}$ using their trust models. However, in some cases the participants may be engaged in repeated interactions that can be exploited by their trust models in order to build accurate trust values of their counterparts. In such situations, the introduction of multiple rounds can compromise the properties of the mechanism by allowing for a greater range of strategies (e.g., cornering the market by consistently offering low prices in initial rounds or accepting losses in initial rounds by providing false and damaging information about competitors). However, the explosion in the strategy space also implies that agents might not be able to compute their optimal strategy due to the intractability of such a process. Now, one way of solving this problem is to constrain the strategies of the agents to be myopic (i.e., best response to the current round) as shown by Parkes and Ungar (2000) using proxy bidding. Another is to allow the agents to learn the trust models without participating in the allocation problem. Then, once the agents have an accurate representation of the trust functions and POS values, the mechanism can be implemented as a one-shot encounter. Note that this problem arises in *any* one-shot mechanism which is implemented in an iterative context and is not solely in the realm of the GTBM.

- **Computational Cost**: As discussed in Section 5, the algorithms we developed to compute the efficient allocation have to be run multiple times to compute the individual payments to the agents for TBMs. Hence, the time needed to compute the allocation and pay the agents may be impractical if the agents have a very limited time to find a solution, put forward a large number of bids, or ask for a large number of tasks to be performed. Hence, it is important that either less complex mechanisms such as those described by Nisan and Ronen (2007) or approximate (and computationally less expensive) algorithms be developed to solve such problems (Archer, Papadimitriou, Talwar, & Tardos, 2003). This will require more work in developing local approximation algorithms and the approximate mechanisms that preserve some of the properties we seek. In this vein, this paper provides a point of departure for these future mechanisms since it provides the efficient mechanisms against which the approximate ones can be compared.

### Acknowledgments

We thank the anonymous reviewers for their highly valuable comments; they have allowed us to improve upon the previous version of this paper, which had a more restrictive mechanism, and also helped rework the proofs. We are grateful to Juuso Välimäki for initial comments on the mechanism, and Ioannis Vetsikas, Enrico Gerding, and Archie Chapman for checking the proofs and discussing the ideas. Juan A. Rodriguez-Aguilar thanks IEA (TIN2006-15662-C02-01), Agreement Technologies (CONSOLIDER CSD2007-0022, INGENIO 2010) and the Jose Castillejo programme (JC2008-00337) of the Spanish Ministry of Science and Innovation. Andrea Giovannucci is funded by a Juan De La Cierva Contract (JCI-2008-03006) and by the EU funded Synthetic Forager project (ICT-217148-SF). Claudio Mezzetti thanks the Fondazione Cassa di Risparmio di Padova e Rovigo for support. The research in this paper was also undertaken as part of the ALADDIN (Autonomous Learning Agents for Decentralised Data and Information Systems) project and is jointly funded by a BAE Systems and EPSRC (Engineering and Physical Research Council) strategic partnership (EP/C548051/1).

## Appendix A. Analysing the Performance of the IP Solution

In this section we analyse the computational performance of the Integer Programming solution we detailed in Section 5 in order to gauge the sizes of problems that can be solved in reasonable time. To this end, it is important to recall that (as was shown in Section 5) the number of input variables to the optimization problem is nearly equal to the number of valuation hyperedges $|\mathcal{E}_v|$, since $|\mathcal{E}_c| \ll |\mathcal{E}_v|$. Given this, we can assume that the performance of the solver is directly related to the number of possible allocations approximated as $|\mathcal{E}_v|$.

Therefore, our test set is composed of several instances of the GTBM Task Allocation Problem characterised by the number of possible allocations. In more detail, to produce such allocations, bids and valuations are generated so that the number of bids submitted by a single bidder and the number of valuations submitted by a single requester follow a geometric distribution with the $p$ parameter set to $0.23$ (Milton & Arnold, 1998) (in order to randomly generate relatively large numbers of bids/asks per agent).[17] A medium-sized problem is set as follows. The number of negotiated tasks is set to 5. The number of task performers is set to 15 and the number of task requesters is set to 20.

---

17. Setting $p$ higher would result in fewer bids/asks per agent.

The average number of generated valuations for each instance is 88 and the average number of bids is 65. Finally, the number of runs of the experiments is 300. Our experiments were performed on a Xeon dual processor machine with 3Ghz CPUs, 2 GB RAM and the commercial software employed to solve the Integer Program is ILOG CPLEX 9.1.

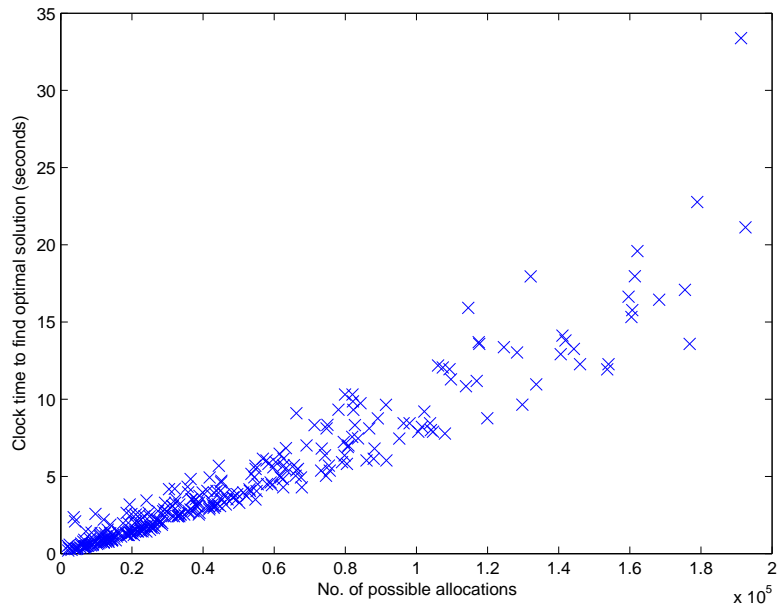

*Figure 2:* Performance of the IP solution.

The results are shown in Figure 2. Specifically, the $x$-axis represents the number of allocations of a given problem instance and the $y$-axis represents the time in seconds elapsed in solving the corresponding problem instance. Notice that the dependence of the difficulty of the problem on the number of allocations is quite clear. Moreover, as can be seen, it is possible to solve a problem with less than $2 \times 10^5$ variables within 40 seconds. It is important to note that the performance of the solver used is critical in this case and future advancements to Mixed Integer Programming (MIP) solvers and CPU clock speeds can only improve our results.

Given these results and since we provide a general formula (see Equation (30)) to compute a priori the number of generated allocations, it is possible to estimate the feasibility of a general problem before performing it. This means that the system designer can ask task requesters and performers to constrain the number of tasks they ask for or the number of bids they issue to come up with an input that can be solved by the program in a reasonable time. It will be more important, however, to design special purpose algorithms that can deal with larger inputs and this is left as future work.